\documentstyle[prb,preprint,aps]{revtex}
\begin{document}
\draft
\tightenlines

\title{Double-layer Heisenberg antiferromagnet at finite temperature:
Brueckner theory and quantum Monte Carlo simulations}

\author{P. V. Shevchenko,$^a$ A. W. Sandvik,$^b$ and O. P. Sushkov~$^a$}

\address{
$^a$ School of Physics, The University of New South Wales,
Sydney 2052, Australia\\
$^b$ Department of Physics, University of Illinois at Urbana-Champaign,
Urbana, Illinois 61801} 

\date{\today}

\maketitle

\begin{abstract}
The double-layer Heisenberg antiferromagnet with intra- and inter-layer
couplings $J$ and $J_\perp$ exhibits a zero temperature quantum 
phase transition between a quantum disordered dimer phase for $g>g_c$ 
and a Neel phase with long range antiferromagnetic order for $g<g_c$,
where $g=J_\perp/J$ and $g_c \approx 2.5$. We consider the
behavior of the system at finite temperature for $g \ge g_c$ using two 
different and complementary approaches; an analytical Brueckner approximation
and numerically exact quantum Monte Carlo simulations. We calculate
the temperature dependent spin excitation spectrum (including the triplet
gap), dynamic and static structure factors, the specific heat, and the
uniform magnetic susceptibility. The agreement between the analytical
and numerical approaches is excellent. For $T \to 0$ and $g \to g_c$, our 
analytical results for the specific heat and the magnetic susceptibility 
coincide with those previously obtained within the nonlinear $\sigma$ 
model approach for $N\rightarrow \infty$. Our quantum Monte Carlo 
simulations extend to significantly lower temperatures than previously, 
allowing us to obtain accurate results for the asymptotic quantum 
critical behavior. We also obtain an improved estimate for the
critical coupling: $g_c = 2.525 \pm 0.002$.

\end{abstract}
\vskip5mm

\vfill\eject

\section{Introduction}

The Heisenberg antiferromagnet on a double-layer square lattice has
become an important model for studying quantum antiferromagnetism in
two dimensions. It was introduced by Millis and Monien as a phenomenological
model to capture the spin-gap behavior observed in high-$T_c$ compounds 
such as YBa$_{\rm 2}$Cu$_{\rm 3}$O$_{\rm 6+x}$,\cite{Millis,Millis1} in 
which the basic structural unit is a pair of CuO$_{\rm 2}$ planes. 
The Heisenberg bilayer with inter- and intra-layer couplings 
$J$ and $J_\perp$ can be tuned through an antiferromagnetic 
order-disorder transition at zero temperature by varying the ratio 
$g = J_\perp/J$. For $g < g_c$ there is long-range 
antiferromagnetic order at zero temperature,
whereas for $g > g_c$ the tendency to singlet 
fomation across the layers leads to a disordered ground state and a spin gap. 
The critical ratio $g_c \approx 2.5$ has been determined using a variety 
of numerical methods.\cite{Hida1,Sandvik,Weihong} The inter-layer coupling 
in bilayer cuprates is typically much smaller than this value; neutron 
scattering experiments indicate $g \approx 0.1$ in 
YBa$_{\rm 2}$Cu$_{\rm 3}$O$_{\rm 6}$.\cite{gexp} Although experimental 
realizations of antiferromagnetic bilayers with $g \approx g_c$ 
are currently lacking, the model in this regime is important for 
theoretical investigations of quantum critical and quantum disordered 
behavior.\cite{Sachdev} 

It has been argued by Haldane \cite{Haldane} and Chakravarty, Halperin and
Nelson \cite{Chakravarty} that the $T=0$ quantum phase transition in 
two-dimensional quantum antiferromagnets is described by the 2+1-dimensional 
nonlinear $O(3)$ $\sigma$-model, and therefore the universality class should 
be that of the 3D classical Heisenberg model. The nonlinear $\sigma$-model 
has three distinct regimes in the $T-g$ plane near the quantum critical
point. For $g<g_{c}$ the long-range antiferromagnetic order at $T=0$ 
is destroyed by thermal fluctuations for any $T > 0$. At low temperatures, 
in the so-called renormalized classical regime, the correlation length 
diverges exponentially as $T \to 0$; $\xi \propto e^{2\pi\rho_s/T}$, 
where $\rho_s$ is the spin stiffness. For $g>g_c$ the singlet-triplet 
excitation gap implies a low-temperature quantum disordered regime 
where the correlation length is temperature independent and remains 
finite as $T\to 0$. Exactly at $g_c$, $\rho_s$ vanishes and the 
correlation length diverges as $1/T$. This is the leading behavior 
also in the high-temperature quantum critical regime for 
both $g < g_c$ and $g > g_c$. Exactly at $g=g_c$, the quantum
critical regime extends down to $T=0$, whereas for $g\neq g_c$ there 
is a crossover to either renormalized classical or quantum 
disordered behavior at lower temperature.

The mapping of the Heisenberg model to the nonlinear $\sigma$-model has been 
rigorously proven only for the single-layer square lattice antiferromagnet 
with nearest-neighbor interactions,\cite{Haldane} which itself does not 
exhibit a zero-temperature quantum phase transition. Assuming the 
$\sigma$-model description, the universal dynamic and static properties 
of two-dimensional antiferromagnets in the vicinity of a zero-temperature 
phase transition have been studied in detail by Sachdev, Ye, and Chubukov 
\cite{Sachdev0,Chubukov0} using an $1/N$ expansion ($N=3$ is the number 
of components of the $\sigma$-model). They found that close to 
criticality, many physical observables such as the specific heat, the
magnetic susceptibility, e.t.c., depend  in a universal manner on a 
small number of  model dependent parameters. The Heisenberg bilayer 
is an ideal model to numerically test these predictions.\cite{Sandvik} 
The model is not frustrated, and therefore the ``sign problem'' hampering
simulations of the single-layer $J_1$-$J_2$ model is not present. 
Compared to other dimerized models, the bilayer has the advantage 
that the square lattice symmetry in the planes is not broken, and 
therefore the asymptotic behavior should be more easily 
observed.\cite{awsmv}

The zero-temperature transition of the bilayer system has been well studied 
by various numerical methods, including dimer-series expansions by Hida
\cite{Hida1} and Zheng, \cite{Weihong} and quantum Monte Carlo simulations 
by Sandvik and Scalapino.\cite{Sandvik} These studies determined the 
critical point in the range be $g_c\approx 2.50-2.56$. The scaling behavior
is consistent with 3D Heisenberg critical exponents.

The finite temperature properties have been also extensively studied by 
numerical methods. Quantum Monte Carlo simulations were performed by Sandvik 
and collaborators \cite{Sandvik,Sandvik1} and showed a reasonably 
good agreement with the predicted quantum critical behavior. High-temperature 
expansions were carried out by Oitmaa {\it et al.}\cite{Oitmaa} and 
finite-cluster 
exact diagonalisation was used by Jaklic and Prelovsek.\cite{Jaklic} Finally, 
high-order strong-coupling thermodynamic expansions for this model
were developed by Elstner and Singh.\cite{Singh1} Unfortunately, all
the methods have so far been restricted to relatively high temperatures. 
A particular recent concern is that the strong-coupling expansions by 
Elstner and Singh indicate a change in the behavior of the susceptibility 
and the specific heat close to the lowest temperature considered in 
the Monte Carlo simulations.\cite{Singh1} It is not clear whether this 
is due to the break-down of the combined high-temperature and strong-coupling 
expansion, which is expected at low temperature, or if the Monte 
Carlo results do not reflect the true asymptotic quantum critical regime.

On the analytical side, until recently the progress with the bilayer
model was less impressive because the conventional spin-wave theory used 
by Hida \cite{Hida1} and the Schwinger boson mean field approach applied by 
Millis and Monien \cite{Millis} were not able to take into account in 
an appropriate way the strong interactions between elementary 
triplets. These approaches yielded results which were inconsistent 
with the numerical studies. In particular, a much too high critical 
coupling $g_c \approx 4.5$ was obtained. A more sophisticated
treatment by Chubukov and Morr,\cite{Morr} taking into account longitudinal 
spin fluctuations which are important close to $g_c$, gave an improved
value $g_c \approx 2.7$ and was also able to explain other
aspects of the quantum Monte Carlo results.

A very efficient analytical approach to deal with a dimerized disordered 
quantum spin system at zero temperature has recently been developed by 
Kotov {\it et al.}.\cite{Sushkov} This method is based on the same 
ideas as the Brueckner theory of nuclear matter,\cite{Brueckner}
and we therefore call it the ``Brueckner method''. This approach has 
already led to rather spectacular success in several problems, including 
a calculation of $g_c$ (yelding $g_c=2.60$), 
the zero temperature critical index, and  
the excitation spectrum for the bilayer model,\cite{Sushkov,Shevchenko}
the elementary excitation spectrum and bound states for one-dimensional 
spin-ladders and spin chains \cite{Sushkov1,Shevchenko1}, as well as 
the ground state structure and the excitation spectrum of the 
frustrated two-dimensional $J_1-J_2$ model.\cite{Kotov}

One of the purposes of the present work is to develop a generalization of the 
Brueckner method to the case of finite temperature, using as a testing 
ground the Heisenberg bilayer Hamiltonian. All analytical results
are general, however, and can be applied to any dimerized quantum spin system, 
including one-dimensional spin ladders and chains. In the present paper 
we focus on the quantum disordered and quantum critical regimes of the
bilayer. Our formulas can be formally extended to the renormalized 
classical regime, but in that case there are very important
corrections that we plan to consider in a future study.

In order to test the reliability of this new theoretical approach, we have 
also carried out quantum Monte Carlo simulations at considerably lower 
temperatures than before. This allows us to definitely conclude that the 
asymptotic quantum critical regime has been reached, and to resolve 
the above mentioned questions arising from strong-coupling expansions. 
The simulations also give an improved estimate of the critical
coupling; $g_c = 2.525 \pm 0.002$.

Using the Brueckner method, we have calculated the temperature dependent 
spectrum of excitations, dynamic and static structure factors and
susceptibilities, the specific heat, and the uniform magnetic 
susceptibility. In the limit $T \to 0$, $g \to g_c$, our results for 
the correlation length, the specific heat, and the magnetic susceptibility 
coincide with those previously obtained using $N\rightarrow \infty$ 
calculations for the nonlinear $\sigma$-model.\cite{Sachdev0,Chubukov0} 
The Monte Carlo approach gives numerically exact results for the 
thermodynamics. The agreement with the analytical calculations is excellent
at low temperature ($T/J \alt 0.5$). Using sum rules, we can also extract an 
approximate excitation spectrum from the quantum Monte Carlo results. 
Also in this case the agreement with the theory is very good.

The rest of the paper is organized as follows. In Sec.~II we first  summarize
the main steps of the Brueckner approach at zero temperature, and then 
proceed to the new calculations at finite temperature. In Sec.~III we 
discuss the Monte Carlo simulations and present data allowing us to extract an 
accurate value for $g_c$. We also discuss finite-size effects. We present 
comparisons between analytical and numerical results for various
physical quantities in Sec.~IV. In Sec.~V we summarize and discuss 
our main conclusions.

\section{Brueckner Theory}

The double-layer Heisenberg model we study is defined by the Hamiltonian
\begin{equation}
\label{H}
H=J\sum_{\langle i,j\rangle}({\bf S}_{1i} \cdot {\bf S}_{1j}+
{\bf S}_{2i}\cdot {\bf S}_{2j})
+J_{\perp}\sum_{i}{\bf S}_{1i} \cdot {\bf S}_{2i},
\end{equation}
where $S_{a,i}$ is a spin-$1/2$ operator at site $i$ of layer $a$
($a=1,2$),  $\langle i,j\rangle$ denotes a pair of nearest-neighbor 
sites on the square lattice, and the total number of sites in a layer is 
$L^2$. Both the coupling constants are antiferromagnetic,
i.e., $J,J_{\perp} >0$. In Sec.~II-A we summarize the main 
steps of the Brueckner diagrammatic approach developed in 
Refs.~\onlinecite{Sushkov,Shevchenko} for this model at zero 
temperature. In Sec.~II-B we generalize the approach to $T > 0$.

\subsection{Zero temperature}

Our approach is formulated in the basis of bond singlets and triplets.
We define a creation operator $t^{\dagger}_{\alpha i}$ for a triplet 
($S=1$) with polarization $\alpha =x,y,z$ at bond $i$, where $i$ 
connects two nearest-neighbor spins ${\bf S}_1(i)$ and ${\bf S}_2(i)$. 
In the bond operator representation \cite{Chubukov1,Sachdev1}
\begin{equation}
\label{repr}
S_{1,2}^\alpha (i)=\frac{1}{2}(\pm t_\alpha\pm t^{\dagger}_\alpha -
i\epsilon_{\alpha \beta \gamma}t^{\dagger}_\beta t_\gamma ),
\end{equation}
and one can exactly map the Hamiltonian (\ref{H}) to the effective Hamiltonian
\begin{equation}
\label{ham}
H_{\rm eff}=H_2+H_4+H_U,
\end{equation}
where
\begin{mathletters}
\begin{eqnarray}
\label{h2}
H_2 & = & \sum_{\alpha , i}J_\perp t^{\dagger}_{\alpha i}t_{\alpha i}+
\frac{\lambda J}{2}(t^{\dagger}_{\alpha i}t_{\alpha i+1}+
t^{\dagger}_{\alpha i}t^{\dagger}_{\alpha i+1}+h.c), \\
\label{h4}
H_4 & = & \frac{J}{2}\sum_{\langle i,j\rangle,\alpha ,\beta}
(t_{\alpha i}^{\dagger}t^{\dagger}_{\beta j} t_{\beta i}t_{\alpha j}-
t^{\dag}_{\alpha i}t^{\dag}_{\alpha j} t_{\beta i}t_{\beta j}), \\
\label{cst}
H_U & = & U\sum_{i,\alpha , \beta}t^{\dagger}_{\alpha i}t^{\dagger}_{\beta i}
t_{\beta i} t_{\alpha i},\hspace{0.4cm} U\rightarrow \infty ,
\end{eqnarray}
\end{mathletters}
Hereafter we set $\lambda=1$, except at the end of Sec.~II-B where 
it will be convenient to use a variable $\lambda$ to formally separate 
the contributions from quadratic and quartic terms in the Hamiltonian.

At the quadratic level the Hamiltonian (\ref{ham}) can be diagonalized 
by a combination of Fourier, 
\begin{equation}
\label{fourier}
t^{\dagger}_{{\bf k},\alpha}=
\frac{1}{L}\sum_{\bf r}t^{\dagger}_{{\bf r},
\alpha}e^{i\bf {(k+k_0) r}},~~~~ {\bf k}_0=(\pi,\pi),
\end{equation}
and Bogoliubov,
\begin{equation}
\label{Bog}
t_{\bf k \alpha}=u_{\bf k}a_{\bf k\alpha}+v_{\bf k}a^{\dagger}_{-\bf k\alpha},
\end{equation}
transformations. This gives the excitation spectrum 
\begin{equation}
\tilde \omega_{\bf k}^2=A_{\bf k}^2-B_{\bf k}^2, 
\end{equation}
where 
\begin{mathletters}
\begin{eqnarray}
A_{\bf k} & = & J_\perp+2J\xi_{\bf k},\\
B_{\bf k} & = & 2J\xi_{\bf k},
\end{eqnarray}
\end{mathletters}
and
\begin{equation}
\xi_{\bf k} = -(\mbox{cos}k_x+\mbox{cos}k_y)/2.
\end{equation}
The momentum takes values within the Brillouin zone $-\pi< k_x,k_y \le \pi$,
but for convenience we have shifted the argument in the Fourier transform 
(\ref{fourier}) by ${\bf k}_0=(\pi,\pi)$. In this notation the minimum 
of the spin-wave dispersion is at ${\bf k}=0$.

An infinite on-site repulsion between triplets, $H_U$, has been introduced
in order to take into account the hard-core constraint 
$t_{\alpha i}^{\dagger}t_{\beta i}^{\dagger}=0$ (only one triplet 
can be excited on a bond). For $g \agt g_c\approx 2.5$, 
the contribution of the quartic
term $H_4$ is relatively small and $H_U$ therefore gives the dominant 
contribution to the renormalization of the spin-wave spectrum. 
Since $H_U$ is infinite the exact scattering amplitude 
\begin{equation}
\Gamma_{\alpha\beta,\gamma\delta}=
\Gamma(\bf K)(\delta_{\alpha\gamma}\delta_{\beta\delta}+
\delta_{\alpha\delta}\delta_{\beta\gamma}),
\end{equation}
where ${\bf K}=({\bf k},\omega)$ is the total energy and momentum of the 
incoming particles, can be found from the following equation, which 
is diagrammatically represented in Figure \ref{fig1}(a),
\begin{equation}
\label{ver}
\Gamma({\bf K})=
-\left(\frac{1}{L^2}\sum_{\bf p}\frac{
u_{\bf p}^2u_{\bf{k-p}}^2}
{\omega-\omega_{\bf p}-\omega_{\bf{k-p}}}\right)^{-1}.
\end{equation}
The Bogoliubov coefficients are 
\begin{equation}
u_{\bf k}^2,v_{\bf k}^2=\pm 1/2+A_{\bf k}/(2\omega_{\bf k}).
\end{equation}
The basic approximation made in the derivation of $\Gamma({\bf K})$ is 
the neglect of all anomalous scattering vertices which are present in 
the theory due to existence of the anomalous Green's function,  
\begin{equation}
G_a({\bf k},t)=-i\langle T(t^{\dagger}_{-{\bf k}\alpha}(t)
t^{\dagger}_{{\bf k}\alpha}(0))\rangle.
\label{ganom}
\end{equation}
The crucial observation is that all anomalous contributions are suppressed 
by a small parameter of the theory --- the density of triplet excitations,
\begin{equation}
\rho=\langle t_{\alpha i}^{\dagger}t_{\alpha i}\rangle
=\frac{3}{L^2}\sum_{\bf q} 
v_{\bf q}^2. 
\end{equation}
It was found in Ref.~\onlinecite{Sushkov} that $\rho\approx 0.12$ at
the critical point, $g=g_c$, and that $\rho$ decreases as $g$ increases. 

The normal self-energy corresponding to the scattering amplitude 
$\Gamma({\bf K})$ 
is given by the equation shown diagrammatically in Figure \ref{fig1}(b):
\begin{equation}
\label{en}
\Sigma({\bf k},\omega)=\frac{4}{L^2}\sum_{\bf q}v_{\bf q}^2
\Gamma({\bf k+q},\omega-\omega_{\bf q}).         
\end{equation}
In order to find the renormalized spectrum, one has to solve the coupled
Dyson equations for the normal Green's function 
\begin{equation}
G_n({\bf k},t)=
-i\langle T(t_{{\bf k}\alpha}(t)t^{\dagger}_{{\bf k}\alpha}(0))\rangle
\label{gnormal}
\end{equation}
and the anomalous (\ref{ganom}). Further separation into a quasiparticle 
contribution and  incoherent background yields
\begin{equation}
\label{sp}
\omega_{\bf k}=Z_{\bf k}\sqrt{\tilde A_{\bf k}^2-\tilde B_{\bf k}^2},
\end{equation}
where
\begin{mathletters}
\label{ab}
\begin{eqnarray}
&&\tilde A_{\bf k}=J_{\perp}+2J\xi_{\bf k}+\Sigma({\bf k},0)+
4J\xi_{\bf k}\frac{1}{L^2}\sum_{\bf q}\xi_{\bf q}v_{\bf q}^2,\\
&&\tilde B_{\bf k}=2J\xi_{\bf k}-4J\xi_{\bf k}
\frac{1}{L^2}\sum_{\bf q} \xi_{\bf q} u_{\bf q}v_{\bf q}.
\end{eqnarray}
\end{mathletters}
We also define
\begin{equation}
\label{zk}
Z_{\bf k} = \left(1-\left.\frac{\partial \Sigma}{\partial \omega}
\right|_{\omega=0}\right)^{-1},\\
\end{equation}
and
\begin{equation}
\label{uv}
U^2_{\bf k},V^2_{\bf k} = \frac{Z_{\bf k}\tilde A_{\bf k}}
{2\omega_{\bf k}}\pm\frac{1}{2}.\nonumber
\end{equation}
The renormalized coefficients $\tilde A_{\bf k}$ and $\tilde B_{\bf k}$ 
(\ref{ab}) also take into account the quartic interaction (\ref{h4}) 
in the one-loop approximation because the corresponding 
effect is small. In order to find the spectrum, 
equations (\ref{ver}),(\ref{en}), and (\ref{sp})-(\ref{uv})
with the substitutions
\begin{equation}
\label{repl}
u_{\bf k}\rightarrow \sqrt{Z_{\bf k}}U_{\bf k},~~~ v_{\bf k}
\rightarrow \sqrt{Z_k}V_{\bf k}
\label{substuv}
\end{equation}
have to be solved self-consistently for $\Sigma({\bf k},0)$ and 
$Z_{\bf k}$. 

The results of the self-consistent solution for the spin-wave gap, 
$\Delta_0=\omega_{k=0}$, are in excellent agreement with a dimer 
series expansions \cite{Sushkov} even close to the critical point, 
as shown in Figure \ref{fig2}. The value of the critical coupling
is $g_c\approx 2.60$, only slightly larger than the numerically determined
value $g_c = 2.51 \pm 0.02$.\cite{Sandvik} It has been demonstrated 
in Ref.~\onlinecite{Shevchenko} that the Brueckner approximation itself 
results in a critical behavior for the triplet gap, 
$\Delta_0\sim (g-g_c)^{\nu}$, with critical index $\nu=1/2$. 
It was also shown that for the model under consideration the actual 
small parameter is $\rho\mbox{ln}(J/\Delta_0)$. Thus when 
the gap is very small the approach can fail. Nevertheless, the critical 
behavior of the gap can be analyzed analytically 
(see Ref.~\onlinecite{Shevchenko}) by 
expanding the Brueckner equations in powers of the triplet density $\rho$.
In the leading approximation the contribution from the quasiparticle 
residue to the renormalization (\ref{repl}) can be neglected and 
we have to set $u_{\bf k}^2=1$ in the vertex (\ref{ver}): 
\begin{equation}
\label{ver1}
\Gamma({\bf q},-\omega_{\bf q})=\left(
\int\frac{d^2{\bf p}}{(2\pi)^2}\frac{1}{\omega_{\bf q}
+\omega_{\bf p}+\omega_{\bf {p-q}}}\right)^{-1}.
\end{equation}
Then the variation of the self-energy (\ref{en}) with the deviation 
$\delta g=g-g_{c}$ results in a linear relation between gap, 
$\Delta_0$, and $\delta g$ near the critical point:
\begin{equation}
\label{lin}
\Delta_0/J\approx a(g-g_{c}),
\end{equation}
where the  constant $a\approx 1.1$ has been evaluated in 
Ref.~\onlinecite{Shevchenko}. The critical exponent $\nu=1$ 
corresponding to (\ref{lin}) is the same as the one predicted within 
the nonlinear $\sigma$ model for $N\rightarrow \infty$.\cite{Chubukov0} 
The plot of the gap, $\Delta_0(g)$, obtained from the self-consistent 
solution of low-density Bruekner equations, i.e., the solution of 
Eqs.~(\ref{en}) and (\ref{sp})-(\ref{uv}) together with the vertex 
(\ref{ver1}) [and the substitutions 
$u_{\bf k},v_{\bf k}\rightarrow U_{\bf k},V_{\bf k}$] is also 
shown in Figure \ref{fig2}. The resulting value of the critical coupling, 
$g_c\approx 2.32$, is about $10$\% lower than numerically determined one.

Consideration of the first correction due to the finite triplet density,
i.e. using $u^2_{\bf p}u^2_{\bf{k-p}} =
(1+v_{\bf p}^2)(1+v^2_{\bf{k-p}}) 
\approx 1+v_{\bf p}^2+v_{\bf {k-p}}^2$ in the vertex (\ref{ver}), 
results in a logarithmic correction at small momenta,
$\Gamma(q,-\omega_q)\approx\Gamma_c(1+\beta\mbox{ln}q)$,
and yields
\begin{equation}
\Delta_0/J=a\delta g(1-\beta\mbox{ln}\delta g),
\end{equation}
with the  constant $\beta\approx 0.3$  evaluated in 
Ref.~\onlinecite{Shevchenko}
Assuming scaling behavior, $\Delta_0\sim (\delta g)^\nu$, the logarithmic 
correction results in a critical index, $\nu=1-\beta\approx 0.67\pm 0.03$, 
which agrees with numerical studies and the prediction of the
$\sigma$ model ($\nu \approx 0.70$). This concludes the solution 
of the two-layer Heisenberg antiferromagnet at zero temperature.

\subsection{Finite temperature}

Having described in the previous section the Brueckner method at
zero temperature as a starting point, we next consider the bilayer Heisenberg 
antiferromagnet at finite temperature. The generalization of the 
method to $T>0$ is simple because in essence it is a 
low-density approximation. Thus it is convenient to use the Feynman 
technique instead of the Matsubara one. In this approximation 
they are equivalent to each other.

Only the normal Green's function
$G_n({\bf k},t)=-i
\langle T(a_{{\bf k}\alpha}(t)a^{\dagger}_{{\bf k}\alpha}(0))\rangle$
exists for physical operators $a_{{\bf k}\alpha}$. At finite 
temperature in an idial gas approximation, it can be written via retarded and advanced functions, 
$G^{R,A}({\bf k},\omega)=[\omega-\omega_{\bf k}\pm i0]^{-1}$,
as (see Ref.~\onlinecite{Abrikosov})
\begin{equation}
G_n({\bf k},\omega)=\frac{1}{2}[G^R+G^A]+
\frac{1}{2}\mbox{ctanh}\frac{\omega}{2T}[G^R-G^A].
\end{equation}
Then, after Fourier (\ref{fourier}) and Bogoliubov (\ref{Bog}) transformations the 
normal and anomalous Green's functions
\begin{mathletters}
\begin{eqnarray}
G^T_n({\bf k},t) & = &
-i\langle T(t_{k\alpha}(t)t^{\dagger}_{k\alpha}(0))\rangle, \\ 
G^T_a({\bf k}, t)& = &
-i\langle T(t^{\dagger}_{-k\alpha}(t) t^{\dagger}_{k\alpha}(0))\rangle,
\end{eqnarray}
\end{mathletters}
at finite temperature are
\begin{equation}
\label{Gn}
G^T_n({\bf k},\omega)=\frac{u_{\bf k}^2(1+n_{\bf k})}{\omega-\omega_{\bf k}
+i0}-\frac{u_{\bf k}^2n_{\bf k}}{\omega-\omega_{\bf k}-i0}-\frac{v_{\bf k}^2
(1+n_{\bf k})}{\omega+\omega_{\bf k}
-i0}+\frac{v_{\bf k}^2n_{\bf k}}{\omega+\omega_{\bf k}+i0},
\end{equation}
\begin{equation}
\label{Ga}
G^T_a({\bf k},\omega)=u_{\bf k}v_{\bf k}
\left(\frac{1+n_{\bf k}}{\omega-\omega_{\bf k}+i0}-
\frac{n_{\bf k}}{\omega-
\omega_{\bf k}-i0}+\frac{n_{\bf k}}{\omega+\omega_{\bf k}+i0}-
\frac{1+n_{\bf k}}{\omega+\omega_{\bf k}-i0}\right),
\end{equation}
where $n_{\bf k}=
\langle
a_{\bf k,\alpha}^+a_{\bf k,\alpha}\rangle
=(e^{\omega_{\bf k}/T}-1)^{-1}$ 
with no summation over $\alpha$. The essential feature is that despite the 
bosonic commutation relations of the single triplet operators, these do 
not have conventional bosonic distribution function because of the strong 
interactions. Actually, $\omega_{\bf k}$ is a functional of $n_{\bf k}$, and 
thus the expression  $n_{\bf k}=(e^{\omega_{\bf k}/T}-1)^{-1}$ 
should be considered as an implicit definition of the thermal distribution 
function $n_{\bf k}$. The density of triplet excitations, trivially 
obtained from the normal Green's function (\ref{Gn}) at finite 
temperature, is given by
\begin{eqnarray}
\label{ni}
\rho^T & = & \langle t^{\dagger}_{i\alpha}t_{i\alpha}\rangle =
i\cdot \lim_{\tau\rightarrow -0}\frac{3}{L^2}\sum_{\bf q}
\int G_n^T({\bf k},\omega)e^{-i\omega \tau}\frac{d\omega}{2\pi} \nonumber \\
& = &\frac{3}{L^2}\sum_{\bf q}(v_{\bf q}^2(1+n_{\bf q})+u_{\bf q}^2 n_{\bf q}).
\end{eqnarray}
Our approach is valid for low density of triplet excitations, and therefore
our consideration is restricted to low temperatures; $T<J_\perp,J$.
We will take into account only leading temperature corrections to the self energy (\ref{en}) neglecting temperature corrections to the vertex (\ref{ver1}).
The quantum transition disappears at any finite temperature and there 
is only a disordered phase with a finite gap towards to triplet 
excitations. Thus we expect that the low-density Brueckner equations 
is an appropriate tool for the model in the $g\ge g_c$ region, 
where the neglected logarithmic corrections, 
$\rho\mbox{ln}(\Delta_T/J)$ and 
$\rho\mbox{ln}(\Delta_0/J)$ are not large and
the vertex in the form (\ref{ver1}) can be used.
Using the 
normal time-dependant Green's function (\ref{Gn}) for the diagram yields
\begin{equation}
\label{ent}
\Sigma^{T}({\bf k},\omega)=\frac{4}{L^2}\sum_{\bf q}
\left[ v_{\bf q}^2(1+n_{\bf q})
\Gamma({\bf k+q},\omega-\omega_{\bf q})+
u_{\bf q}^2n_{\bf q}\Gamma({\bf k+q},\omega+\omega_{\bf q})\right].
\end{equation}
In order to find the renormalized spectrum at finite temperature, 
one has to solve the coupled Dyson equations for the normal and anomalous 
Green's functions (\ref{Gn}) and (\ref{Ga}), as was done in Sec.~II-A at 
zero temperature. Further separation into a quasiparticle contribution 
and incoherent background renormalizes the coefficients 
\begin{mathletters}
\label{abt}
\begin{eqnarray}
\tilde A_{\bf k} & = &
J_\perp+2J\xi_{\bf k}+\Sigma^T({\bf k},0)
+4J\xi_{\bf k}\frac{1}{L^2}\sum_{\bf q}\xi_{\bf q}
[v_{\bf q}^2(1+n_{\bf q})+u_{\bf q}^2n_{\bf q}],\\
\tilde B_{\bf k} & = & 
2J\xi_{\bf k}
-4J\xi_{\bf k}\frac{1}{L^2}\sum_{\bf q}\xi_{\bf q}
u_{\bf q}v_{\bf q}(1+2n_{\bf q}),
\end{eqnarray}
\end{mathletters}
where the effect due to $H_4$ is also taken into account in the the one-loop approximation.
Then the finite temperature renormalized spectrum $\omega_{\bf k}$, 
the renormalization constant $Z_{\bf k}$, and the Bogoliubov coefficients 
$U_{\bf k}$ and $V_{\bf k}$ are given by Eqs.~(\ref{sp}) and (\ref{uv}) 
with the self-energy (\ref{ent}) and the coefficients (\ref{abt}).

In the low-density approximation the 
equations (\ref{sp}),(\ref{zk}),(\ref{uv}),(\ref{ver1}),(\ref{ent}), 
and (\ref{abt}) should be solved self-consistently for 
$\Sigma^T({\bf k},0)$ and $Z({\bf k})$ with the substitution
$u_{\bf k}\rightarrow U_{\bf k}$, $v_{\bf k}\rightarrow V_{\bf k}$.

All thermodynamic functions of the system can be obtained from 
free energy ${\bf F}$. The total energy of the system can 
not be found as a simple summation of quasiparticle energies because 
of the strong interactions, but it can be calculated in following way:
Assume that the  parameter $\lambda$ in the quadratic term (\ref{h2}) 
takes values between $0$ and $1$. Then we have
\begin{equation}
\label{Fl}
\frac{\partial {\bf F}}{\partial \lambda}=\left\langle\frac{\partial H}
{\partial \lambda} \right\rangle_T.
\end{equation}
If $\lambda=0$, the system is a set of dimers with free energy
\begin{equation}
{\bf F}_0=-T\mbox{ln}(1+3e^{-J_{\perp}/T}).
\end{equation}
Integrating Eq.~(\ref{Fl}) then yields
\begin{equation}
\label{F}
{\bf F}-{\bf F}_0=\int_0^{1}d\lambda
\left\langle\frac{\partial H}{\partial \lambda}\right\rangle_T,
\end{equation}
where
\begin{eqnarray}
\label{F1}
\left\langle\frac{\partial H}{\partial \lambda}\right\rangle_T & = &
\frac{J}{2L^2}\sum_{i\alpha}\langle t^{\dagger}_{\alpha i}
t_{\alpha i+1}+t^{\dagger}_{\alpha i}t^{\dagger}_{\alpha i+1}
+ {\rm h.c.}\rangle 
\nonumber \\
& = & J\frac{6}{L^2}\sum_{\bf k}\xi_{\bf k} (v_{\bf k}^2
[1+n_{\bf k}]+u_{\bf k}^2n_{\bf k}+u_{\bf k}v_{\bf k}[1+2n_{\bf k}]).
\end{eqnarray}
Using Eq.~(\ref{F}), one can calculate the
thermodynamic potential and all thermodynamic functions using the
self-consistent solution for the spectrum $\omega_{\bf k}$ and the
Bogoliubov coefficients $v^2_{\bf k}$ and $u_{\bf k}^2$ 
at finite temperature. 

This concludes the solution of the two-layer Heisenberg antiferromagnet 
model at finite temperature. The results of self-consistent solution of 
the  Brueckner equations for the triplet gap, the spectrum, as well 
as various thermodynamic quantities will be presented in Sec.~IV, along 
with comparisons with numerically exact results obtained using the
quantum Monte Carlo method discussed in the next section.

\section{Quantum Monte Carlo}

Quantum Monte Carlo simulations of the Heisenberg bilayer have previously
been used to determine the $T=0$ critical coupling, with the result
$g=2.51 \pm 0.02$.\cite{Sandvik} This value is also in good agreement 
with results obtained using dimer series expansion methods.
\cite{Hida1,Weihong} Finite-temperature simulations in the quantum 
critical regime \cite{Sandvik,Sandvik1} have shown generally good 
agreement with predictions of $1/N$ calculations for the nonlinear 
$\sigma$ model. However, the algorithm used in the previous simulations 
did not work well at very low temperatures, and therefore there has been some 
concern \cite{Singh1} that the results do not reflect the true 
$T \to 0$ behavior. Recently, the ``stochastic series expansion'' 
algorithm \cite{sse1,sse2} used in previous work has been considerably 
improved by the introduction \cite{sse3} of a new cluster-type 
updating scheme (inspired by the ``loop algorithms'' developed for 
worldline Monte Carlo simulations \cite{loops}) that overcomes the 
limitations of the previous local sampling scheme. We have used this 
algorithm for large-scale calculations in order to obtain a more precise 
estimate of the critical point and to study the low-temperature 
quantum critical and quantum disordered behavior. In this section we 
extract the $T=0$ critical coupling and also discuss effects of finite 
lattice size in $T > 0$ calculations. In the Sec.~IV we will compare 
results for the thermodynamics and the excitation spectrum with the 
Brueckner theory. For details of the simulation algorithm we refer 
to previous literature.\cite{sse2,sse3}

One way to extract the critical point is from the size dependence of
the $T=0$ spin stiffness $\rho_s$. Imposing a uniform twist $\phi$ 
such that the spin-spin interaction is modified according to
\begin{equation}
S^x_{ai}S^x_{aj} + S^y_{ai}S^y_{aj} \to
(S^x_{ai}S^x_{aj} + S^y_{ai}S^y_{aj})\cos{(\phi)} 
+ (S^y_{ai}S^x_{aj} - S^x_{ai}S^y_{aj}) \sin{(\phi)}
\end{equation}
(for all nearest-neighbor spin pairs in either the $x$ or $y$ 
lattice direction in both layers $a=1,2$) the stiffness is given by
\begin{equation}
\rho_s = {\partial ^2 E(\phi) \over \partial^2 \phi },
\end{equation}
where $E(\phi)$ is the internal energy per spin. The stiffness can 
be related to the fluctuation of the ``winding number'' in the 
Monte Carlo simulations \cite{sse2} and can hence be evaluated directly
without actually including the twist. 

We study quadratic lattices with linear size $L$, i.e., the total
number of spins is $2L^2$. According to scaling theory,\cite{wallin} 
at the critical point $\rho_s$ should depend on the system size 
according to
\begin{equation}
\rho_s(q=g_c) \sim L^{d-2-z},
\end{equation}
where $d$ is the dimensionality and $z$ is the dynamic exponent. Hence,
with $z=1$, as follows if the transition is in the universality class of
the $3D$ classical Heisenberg model, $L\rho_s$ graphed versus $g$ for
different system sizes should intersect at $g_c$. In Figure~\ref{fig3}
we show results for $L$ up to $20$, calculated at temperatures
sufficiently low to give the ground state (for the largest system, $L=20$, 
$J/T = 320$ was used). For the largest lattices, the curves 
intersect at $g \approx 2.525$, in good agreement with the previous 
Monte Carlo estimate  $g_c = 2.51 \pm 0.02$ \cite{Sandvik}

Other useful quantities that can be calculated in the simulations
include the static structure factors $S^\pm({\bf q})$ and
susceptibilities $\chi^\pm({\bf q})$, which are defined according to
\begin{mathletters}
\label{sxqdef}
\begin{eqnarray}
S^{\pm}({\bf q}) & = & {1 \over 2L^2}
\sum\limits_{{\bf r}_1,{\bf r}_2} 
{\rm e}^{i{\bf q}\cdot ({\bf r}_2 - {\bf r}_1))}
[S_{1,r_1} \pm S_{2,r_1}][S_{1,r_2} \pm S_{2,r_2}] \\
\chi^{\pm}({\bf q}) & = & {1 \over 2L^2}
\sum\limits_{{\bf r}_1,{\bf r}_2} 
{\rm e}^{i{\bf q}\cdot ({\bf r}_2 - {\bf r}_1))}
\int_0^\beta d\tau
[S_{1,r_1}(\tau) \pm S_{2,r_1}(\tau)][S_{1,r_2}(0) \pm S_{2,r_2}(0)].
\end{eqnarray}
\end{mathletters}

We first consider the uniform magnetic susceptibility, 
$\chi=\chi^+(0,0)$, close to the estimated $g_c$. According to 
calculations for the nonlinear $\sigma$ model, $\chi(T)$ should be 
linear in $T$ in the quantum critical regime, with intercept $0$ at 
$g=g_c$. In order to study the behavior at low temperatures, 
lattices sufficiently large to eliminate finite-size effects have 
to be used. In Figure~\ref{fig4} we graph results at $g=2.53$ for 
several system sizes. The results show that in order to obtain results 
reflecting the thermodynamic limit down to $T/J=0.1$, systems with 
$L=64-128$ are required. Note that the asymptotic $T\to 0$ behavior 
for a finite lattice is always an exponential decay to zero, reflecting 
the finite-size singlet-triplet gap (which scales to zero with 
increasing $L$ for $g \le g_c$ and remains finite for $g > g_c$). 
However, in the temperature regime shown in Fig.~\ref{fig4}, the 
finite-size effects actually enhance $\chi$.

In Figure~\ref{fig5} we show results for system sizes sufficiently large
to eliminate finite-size effects for $g=2.52$ and $g=2.53$. For both
couplings, the behavior is very close to linear for  
$0.1 \le T/J \alt 0.4$.
The intercept of a line fitted to $L=128$ data 
for $T/J \le 0.17$ is
positive for $g=2.52$ and negative for $g=2.53$, in consistency with 
$g_c \approx 2.525$ estimated from the $T=0$ stiffness scaling. Using the 
data shown in Fig.~\ref{fig5} we estimate $g_c \approx 2.525 \pm 0.002$.
This result is slightly lower than estimates ($g_c=2.54-2.56$) from 
dimer series expansions.\cite{Hida1,Weihong}

In Sec.~IV we will make use of sum rules and the ratio
\begin{equation}
R^{\pm}({\bf q}) = {S^\pm({\bf q}) \over T \chi ^\pm({\bf q})}
\end{equation}
to extract the approximate finite-temperature triplet excitation spectrum.
Here, in Figure~\ref{fig6}, we show the finite-size effects in
$R^-(\pi,\pi)$ for $g=2.52$ and $2.53$. The classical value of
$R^\pm({\bf q})=1$ for any ${\bf q}$, and this is also expected to 
hold for $R^-(\pi,\pi)$ of an antiferromagnet in the renormalized 
classical regime (where both the staggered structure factor and
the staggered susceptibility diverge). For a system with a gap, on 
the other hand, both $S^\pm ({\bf q})$ and $\chi^\pm ({\bf q})$ 
converge to finite values for all ${\bf q}$ and $R^-(\pi,\pi)$ 
therefore diverges as $1/T$ at low temperature. Nonlinear $\sigma$-model 
calculations have predicted a temperature independent $R^-(\pi,\pi)$
in the quantum critical regime, with the  value
$R^-(\pi,\pi) \approx 1.09$.\cite{Chubukov0,Sachdev0} Previous Monte 
Carlo calculations showed a reasonable agreement with this prediction close 
to the critical point for $T/J \approx 0.3$.\cite{Sandvik1} The results 
shown in Fig.~\ref{fig6} demonstrate that the ratio is very sensitive 
to the system size at lower temperatures. The finite singlet-triplet gap
present for any finite $L$ causes a $1/T$ behavior of $R$ as $T \to 0$,
starting at a temperature that decreases with increasing $L$. The
results shown in Fig.~\ref{fig6} indicate that $L=128$ should be 
sufficiently large to eliminate finite-size effects for all $T/J \ge 0.1$.
For this lattice size, the data for $g=2.52$ decreases with decreasing 
$T$ whereas there is an increase for $g=2.53$. This behavior is consistent 
with the predicted temperature independence exactly at $g_c$ and 
$g_c \approx 2.525$. In Sec.~IV we will present analytical results 
for the temperature corrections to $R^-(\pi,\pi)$ and make further 
comparisons with numerical results.

\section{Results}

We now discuss finite-temperature calculations at the critical point and in 
the quantum disordered regime. Results of the low-density Brueckner
theory will be compared with quantum Monte Carlo in the $g\geq g_c$ region. 
Note that the critical point resulting from the self-consistent solution 
of the low-density Brueckner equations at zero temperature, 
$g_c\approx 2.320$ (see Figure 2), is different from the actual value 
$g_c\approx 2.525$. Unless stated otherwise, we will consider numerical 
results at $g_c$ as the average of data for $g=2.52$ and $2.53$, which 
should be close to the actual $g_c \approx 2.525$ critical behavior in 
the temperature regime considered. We compare the results with the 
Brueckner theory predictions for the critical point in that approach;
$g=2.320$. To compare our analytical and numerical results in the quantum 
disordered regime we will take $g=3.0$ (where the Brueckner theory 
predictions are in perfect agreement with the numerical data at zero 
temperature) both for the Brueckner and Monte Carlo calculations.

\subsection{Triplet gap}

Let us first consider the critical behavior of the triplet gap, 
$\Delta_T=\omega_{k=0}$, described by the Brueckner equations. 
The temperature and the deviation from the critical point must 
satisfy the condition $T/J,\delta g\ll g_c$. No 
restriction is imposed on the ratio $\delta g/T$, however.

Close to the critical point and for small momenta ($k\ll 1$) the 
dispersion can be represented as
\begin{equation}
\label{o}
\omega_k\approx\sqrt{\Delta_T^2+\gamma^2k^2},
\end{equation}
where $\gamma=1.9J$ is the zero temperature spin-wave
velocity.\cite{Weihong,Sushkov,Hida} Eq.~(\ref{sp}) at the point 
$k=0$ gives
\begin{equation}
\label{d1}
\Delta^2_T=Z_0(\tilde A_0^2-\tilde B_0^2).
\end{equation}
It is convenient to introduce the values of $\tilde A_{0}, \tilde B_{0}$, 
and $Z_0$ at zero temperature $\tilde A_{T=0}, \tilde B_{T=0},  
Z_{T=0}$, which satisfy $\Delta_0^2=Z_{T=0}
(\tilde A_{T=0}^2-\tilde B_{T=0}^2)$. Let us vary the temperature, 
keeping $J_\perp$ and $J$ fixed. The main contribution to the 
variations of the integrals in Eqs.~(\ref{ent}) and (\ref{abt}) comes 
from small momenta; $q\sim \Delta/\gamma\ll 1$. Then we find the 
variations of $\tilde A_k,\tilde B_k$ at $k=0$ to leading order: 
\begin{mathletters}
\label{dab}
\begin{eqnarray}
&&\tilde A_0=\tilde A_{T=0}+\delta \Sigma^T(0,0)-
\frac{J A_cZ_c}{\pi\gamma^2}R(T)\\
&&\tilde B_0=\tilde B_{T=0}+\frac{J A_cZ_c}{\pi\gamma^2}R(T),
\end{eqnarray}
\end{mathletters}
where
\begin{equation}
\label{et}
\delta\Sigma^T(0,0)\approx \frac{\Gamma_c A_cZ_c}{\pi\gamma^2}R(T),
\end{equation}
and
\begin{equation}
R(T)=-\Delta_T+\Delta_0+2T\int_{\frac{\Delta_T}{T}}^{\infty}\frac{dx_1}
{e^{x_1}-1},
\end{equation}
and where $\Gamma_c$, $Z_c$ and $A_c$ denote the vertex $\Gamma(0,0)$, 
the quasiparticle residue $Z_{k=0}$, and $\tilde A_0$ evaluated at 
$T=0$ and  $g=g_c$. 

If we substitute  (\ref{dab}) into (\ref{d1}) and neglect 
terms quadratic in $\Delta_T,\Delta_0$, we find that the variation of 
the self-energy has to vanish; $\delta \Sigma^T(0,0)=0$. Thus  
$R(T)=0$ determines the finite-temperature gap $\Delta_T$ as a 
function of the zero temperature gap $\Delta_0$:
\begin{equation}
\label{gap}
\frac{\Delta_T}{T}\approx\frac{\Delta_0}{T}+2\int_{\frac{\Delta_T}{T}}^{\infty}
\frac{dx}{e^x-1}.
\end{equation}
Also, we have found that 
\begin{equation}
\delta \rho^T\approx\frac{3A_cJ}{4\pi\gamma^2}R(T)=0.
\end{equation}
Thus, as the temperature increases the quantum fluctuations are 
reduced while the temperature fluctuations are enhanced, the total 
triplet density remaining constant. After a simple integration 
of (\ref{gap}) we obtain the analytical solution  
\begin{equation}
\label{a}
\Delta_T=Ty(x),
\end{equation}
where
\begin{equation}
y(x)=2\mbox{ln}\left[\frac{e^{x/2}}{2}+\sqrt{1+\frac{e^x}{4}}\right]=2\mbox{arcsh}\left[\frac{1}{2}e^{x/2}\right],
\end{equation}
with $x=\Delta_0/T$.  The variable $x$ determines whether the 
antiferromagnet is in the quantum critical ($x\rightarrow 0$), 
quantum disordered ($x\rightarrow \infty$), or the renormalized 
classical ($x\rightarrow -\infty$) region. It corresponds to the
variables $x_1=-2/x$ and $x_2=1/x$ introduced within the nonlinear
$\sigma$ model formalism \cite{Chubukov0} for the critical 
regions in the $T-g$ plane. The zero temperature gap, 
$\Delta_0/J=a(g- g_c)$, is defined for positive values 
($g>g_c$) but we analytically continue it to negative values ($g<g_c$).
The temperature gap $\Delta_T$ depends on the coupling $g$ via the
zero temperature gap $\Delta_0(g)$. Thus, in order to calculate 
$\Delta_T(g)$ one should substitute Eq.~(\ref{lin}) into (\ref{a}).

The universal function $y(x)$ for the inverse correlation length, 
$1/\xi=\Delta_T$, is exactly the same as that previously obtained for the
nonlinear $\sigma$ model ($N\rightarrow \infty$),\cite{Chubukov0} 
with the limiting behavior
\begin{mathletters}
\label{lim}
\begin{eqnarray} 
y(x) & = & 2\mbox{ln}\frac{1+\sqrt{5}}{2}+\frac{x}{\sqrt{5}},
\hspace{0.3cm} x\rightarrow 0,\\
y(x) & \approx & x+2e^{-x},\hspace{0.3cm} x\rightarrow \infty ,\\
y(x) & \approx & e^{x/2},\hspace{0.3cm} x\rightarrow -\infty .
\end{eqnarray}
\end{mathletters}
Thus the correlation length is exponentially large in the renormalized 
classical region, proportional to $1/T$ in the quantum-critical region, 
and of the order $1/\Delta_0$ in the quantum disordered regime.
At the crititical point ($x=0$) we have
\begin{equation} 
\label{y0}
\Delta_T=y_0T,\hspace{0.3cm} 
y_0=2\mbox{ln}\left[\frac{1+\sqrt{5}}{2}\right]\approx 0.962424.
\end{equation}
In Figure {\ref{fig7} we compare the analytical result (\ref{a}) at $g=g_c$, 
and $g=3.0$  obtained under the assumption of small $T$ and $\delta g$ 
with the corresponding self-consistent numerical solutions of the 
Brueckner equations for the triplet gap. The agreement is very good
for $g=g_c$ at low temperatures, and slightly worse at $g=3.0$.

Now, let us consider the following dimensionless ratio 
\begin{equation}
\label{wx}
R({\bf q})=\frac{S({\bf q})}{T\chi({\bf q})},
\end{equation}
where $S({\bf q})$, $\chi({\bf q})$ are the odd ($-$) static structure factor
and susceptibility defined in Eqs.~(\ref{sxqdef}) [with the momentum shifted
by $(\pi,\pi)$]. They are related to the corresponding dynamic
structure factor, 
\begin{equation}
S({\bf q},\omega)=
\mbox{Im}\int e^{i\omega t}\langle
T({\bf S}({\bf q},t){\bf S}(-{\bf q},0))\rangle dt,
\end{equation}
where ${\bf S}_i={\bf S}_{1i}-{\bf S}_{2i}$,  
via the sum rules\cite{Hohenberg}
\begin{mathletters}
\begin{eqnarray}
S({\bf q}) & = &
\frac{1}{\pi}\int_0^{\infty}S({\bf q},\omega)\left(1+e^{-\omega/T}\right)
d\omega, \label{sumrul1} \\
\chi({\bf q}) & = & \frac{2}{\pi}\int_0^{\infty}S({\bf q},\omega)\left
(1-e^{-\omega/T}\right)\frac{d\omega}{\omega} \label{sumrul2}.
\end{eqnarray}
\end{mathletters} 
Taking into account the contribution from the elementary triplet only, 
 the structure factor can be written in the form 
\begin{equation}
\label{sdelta}
S(q,\omega)=(u_{\bf q}+v_{\bf q})^2\delta(\omega-\omega_{\bf q}).
\end{equation}
Then, using the above sum rules (\ref{sumrul1}) and (\ref{sumrul2}), 
the dimensionless ratio $R({\bf q})$ can be rewritten in the form
\begin{equation}
\label{wq}
R({\bf q})=\frac{1+e^{-\omega_{\bf q}/T}}
{1-e^{-\omega_{\bf q}/T}}\frac{\omega_{\bf q}}{2T}.
\end{equation}
The importance of this ratio is that it can be obtained using Monte Carlo 
simulations by calculating the 
static susceptibility and structure factor, as discussed in Sec.~III.
The spectrum of elementary excitations, $\omega_{\bf q}$, can then be 
extracted from Eq.~(\ref{wq}). At zero momentum, $q=0$, the critical 
behavior of $R(0)$ is given by 
\begin{equation}
\label{wq0}
R(0)=\frac{y(x)}{2}\frac{1+e^{-y(x)}}{1-e^{-y(x)}}.
\end{equation}
Then the limiting behavior of $R(0)$ is readily obtained from Eq.~(\ref{a}):
\begin{mathletters}
\begin{eqnarray}
\label{wx00}
&&R(0)=\frac{\sqrt{5}}{2}(y_0+\frac{x}{\sqrt{5}}(1-\frac{2}{\sqrt{5}}y_0)),
\hspace{0.5cm} x\rightarrow 0,\\
&&R(0) = \frac{x}{2}+e^{-x}(1+x), \hspace{0.5cm} x\rightarrow \infty, \\
&&R(0) = 1 - \frac{e^{x/2}}{2},\hspace{0.5cm} x\rightarrow -\infty.
\end{eqnarray}
\end{mathletters}

We have calculated $R(0)$ as a function of temperature using quantum 
Monte Carlo simulations close to the critical point, at $g=2.52$ and 
$2.53$, and using the Brueckner equations at the critical point $(x=0$).
As discussed in Sec.~III (Fig.~\ref{fig4}), $R(0)$ is very sensitive 
to the lattice size at low temperatures. In Figure \ref{fig8} 
we show Monte Carlo results for systems sufficiently large to 
eliminate finite-size effects (linear size $L$ up to $128$). 
In Sec.~III the critical point was determined to
be $g_c = 2.525 \pm 0.002$. Even for $|g-g_c|$ as small as $0.005$,
there are very large differences between the $g-g_c > 0$ and $g-g_c < 0$ 
data. In order to extract the behavior exactly at $g=g_c$, simulations
would have to be carried out for values of $g$ also between $2.52$ and
$2.53$, and the critical point would have to be determined to
even higher accuracy. The available data neverthelsess shows that
the Brueckner theory is very accurate. The analytical curve falls 
between the two numerical data sets at low temperature, and the $T$
dependence at high temperatures is also very well reproduced. In the 
Brueckner theory $R(0)\rightarrow 1.076$ at $g\rightarrow g_c$ 
as $T\rightarrow 0$. The numerical results suggest that the exact
value is slightly higher, but close to $1.08$.

The temperature dependent triplet gap extracted from the
Monte Carlo data for $R(0)$ using Eq.~(\ref{wq}) is presented in 
Figure \ref{fig7}, along with our analytical prediction. The agreement is 
almost perfect at the critical point, while at $g=3.0$ it is not so 
good but still quite reasonable. 

The formula (\ref{wq}) is not exact, due to the neglected contribution of 
the three-particle continuum and broadening. Note that the contribution 
of the two-particle continuum is canceled by symmetry. The three-particle 
continuum is negligible only near the $q=(0,0)$ at the critical point
while it can be important in other cases.

Using Eq.~(\ref{wq}), we can extract the triplet spectrum $\omega_{\bf k}$ from the
Monte Carlo results for $R({\bf q})$. In Figure \ref{fig9}
we compare it with the solution
of the Brueckner equations at $g=g_c$ and  $T=0.5J$. The 
agreement between the Monte Carlo and analytical results is even 
better then might expected (the agreement remains very good also at
lower temperatures). The Brueckner approximation should be 
accurate if the triplet density $\rho$ is low, but in fact it is 
as high as $\rho\sim 0.1$.\cite{Sushkov} The surprisingly good agreement 
justifies Eq.~(\ref{wq}) and the assumption that the dynamic structure factor 
is peaked at $\omega=\omega_k$ [see Eq.~(\ref{sdelta})]. Thus the 
broadening is negligible and the elementary triplet is well defined 
for $g\ge g_c$ at low temperature. However, the broadening can be important for $g<g_c$.

\subsection{Specific heat}

The calculation of the specific heat, $C_V$, is straightforward. The 
energy of the system cannot be written as a simple summation over 
quasiparticle energies, $E\neq \sum \omega_k n_k$, due to the strong 
interaction. However, one can write the energy variation  as 
$\delta E=\sum \omega_k\delta n_k$, in analogy with the Landau theory 
of Fermi liquids. Thus, the specific heat per spin is
\begin{equation}
\label{Cv}
C_V=\frac{3}{2L^2}\sum_{\bf k} \omega_{\bf k} 
\frac{\partial n_{\bf k}}{\partial T}.
\end{equation}
It is convenient to use Eq.~(\ref{Cv}) instead of taking of the second 
derivative, $C_V=-\frac{1}{2}T\partial^2 F/\partial T^2$, using 
Eq.~(\ref{F}). The specific heat calculated from Eq.~(\ref{Cv}) and
the self-consistent solution of the Bruekner equations is shown in
Figure \ref{fig10} 
as a function of $T$ at $g=g_c$ and $g=3.0$. We have also 
calculated the specific heat by fitting a high-order polynomial 
to the internal energy obtained using quantum Monte Carlo simulations. 
As seen in Fig.~\ref{fig10}, 
the agreement between the analytical and numerical 
results is excellent for $T<0.5J$.
The agreement is perfect at the critical point, while at $g=3.0$ it is not so 
good. 
The reason for this can be traced to the fact that the value 
of triplet gap, $\Delta_0\approx 0.78J$, at $T=0, g=3.0$ 
in our analytic method is slightly different from the numerical 
result $\Delta_0\approx 0.75J$ at zero temperature. This 
difference  leads to a slightly different exponent for the specific heat.

In the limit $T\ll J, g\rightarrow g_c$, the expression for 
the specific heat can be found explicitly by combining Eqs.~(\ref{o}), 
(\ref{a}) and (\ref{Cv}): 
\begin{equation}
\label{Cv1}
C_V=\frac{3\zeta(3)T^2}{2\pi \gamma^2}\Psi(\frac{\Delta_0}{T}),
\end{equation}
where the universal function $\Psi$ is given by
\begin{equation}
\Psi(x)=\frac{1}{2\zeta(3)}\int_{y(x)}^{\infty}
\frac{dx_1x_1e^{x_1}}{(e^{x_1}-1)^{2}}\left[x_1^2-y^2(x)+xy(x)
\frac{dy(x)}{dx}\right].
\end{equation}
The coefficient of $\Psi(x)$ in (\ref{Cv1}) has been chosen to be the 
specific heat of a single gapless Bose degree of freedom with dispersion 
$\omega=\gamma k$ in two dimensions. The value of $\Psi(x)$ at 
$T\rightarrow 0$ is thus a measure of the effective number of such 
modes in the ground state. In the renormalized classical limit, it has 
the value $\Psi(-\infty)=3$, while at the critical point $\Psi(0)=
\frac{4}{5}\Psi(-\infty)$. In the quantum disordered limit, 
$x\rightarrow \infty$, we find exponentially small specific heat:
$$ 
C_V\approx \frac{3T^2}{4\pi\gamma^2}x^3e^{-x}, \hspace{0.3cm}x\rightarrow
\infty.
$$
Thus $\Psi(\infty)=0$, corresponding to the absence of any gapless 
degrees of freedom. The limiting behavior of universal function 
$\Psi(x)$ is exactly the same as obtained one for the nonlinear 
$\sigma$-model ($N\rightarrow \infty$).\cite{Chubukov0}

The fact that $\Psi(-\infty)=3$ implies that our approach does not 
describe well the physics for $g < g_c$, where the spin dynamics is 
well described by rotationally averaged spin-wave (Goldstone) 
fluctuations about a Neel-ordered ground state. There are only 
two gapless modes in the renormalized classical regime, while in 
our approach there are three, and the critical and Goldstone 
fluctuations are indistinguishable. The application of the Brueckner 
theory in this phase requires modifications; a nonzero sublattice 
magnetization, i.e., a single particle condensate at zero momentum, 
has to be introduced. 

\subsection{Uniform magnetic susceptibility}

Let us now calculate another important physical observable --- the
magnetic susceptibility. If a uniform magnetic field of strength
${\bf h}$ is applied, then the term
\begin{equation}
\label{h}
H_h=-{\bf h}\cdot \sum_i {\bf S}_i=ih_\alpha e_{\alpha\beta\gamma}\sum_{\bf k} 
t^{\dagger}_{{\bf k}\beta}
 t_{{\bf k}\gamma},
\end{equation}
should be added to the Hamiltonian (\ref{ham}).

Let us first to concentrate on the zero temperature properties.
The second order correction to the total energy of the system due to the
perturbation (\ref{h}) is given by the diagrams shown, in 
Figure \ref{fig11}(a),
\begin{equation}
E^{(2)}\sim h^2\sum_{\bf k}\int\frac{d\omega}{2\pi}\left[
G_n({\bf k},\omega)G_n({\bf k},\omega)-G_a({\bf k},
\omega)G_a({\bf k},\omega)\right].
\end{equation}
where  
$G_n({\bf k},t)=-i
\langle T(t_{{\bf k}\alpha}(t)t^{\dagger}_{{\bf k}\alpha}(0))\rangle$ 
and $G_a({\bf k},t)=-i<T(t^{\dagger}_{-{\bf k}\alpha}(t)t^{\dagger}_{{\bf k}
\alpha}(0))>$ are normal and anomalous Green's functions at zero 
temperature which can be formally obtained by setting $n_{\bf k}\equiv 0$
in the expressions (\ref{Gn}) and (\ref{Ga}). After a simple integration 
over $\omega$, one can show that $E^{(2)}=0$. Thus, the uniform magnetic 
susceptibility per spin, 
\begin{equation}
\chi=-\frac{1}{2} {\partial^2 {\bf F}\over \partial h^2}
\Bigr |_{h=0},
\end{equation}
is zero at $T=0$. It is worth noting that the Bogoliubov transformation 
(\ref{Bog}) does not change the operator of the magnetic moment
$
M_\alpha=-ie_{\alpha\beta\gamma}\sum_{\bf k}t^+_{{\bf k}\beta}t_{{\bf k}
\gamma}
=-ie_{\alpha\beta\gamma}\sum_{\bf k}a^+_{{\bf k}\beta}a_{{\bf k}\gamma}
$.

To calculate the uniform susceptibility at finite temperature, let us 
demonstrate that within the framework of the Brueckner approach the 
perturbation (\ref{h}) splits a three degenerate triplet into excitations 
at $\omega_k$ and $\omega_k \pm \mu \cdot  h$ with magnetic moment 
$\mu=1$ at zero temperature. The correction to the self-energy due to 
the magnetic field ${\bf h}$ is  given by the diagrams (b) and (c) in 
Figure \ref{fig11}:
\begin{equation}
\label{enh}
\delta\Sigma^{h}_{\beta\gamma}({\bf K})=
ih_\alpha e_{\alpha\beta
\gamma}\left(1-i\int\Gamma({\bf K+Q})\left[G_n({\bf Q})G_n({\bf Q})
-G_a({\bf Q})G_a({\bf Q})\right]\frac{d^3{\bf Q}}{(2\pi)^3}-
\right.\end{equation}
$$
-
\left.
5\int \Gamma^2({\bf K+Q})G_n({\bf K+Q-Q_1})
G_n^2({\bf Q}_1)G_n({\bf Q})\frac{d^3{\bf Q}_1}{(2\pi)^3}\frac{d^3
{\bf Q}}{(2\pi)^3}\right)=
$$
$$=
ih_\alpha e_{\alpha\beta\gamma}\left[1-\left.\frac{\partial\Sigma^{h=0}
}{
\partial \omega}\right|_{\omega=0}\right]=ih_\alpha 
e_{\alpha\beta\gamma}Z_{\bf k}^{-1}.$$
The first diagram in Fig.~\ref{fig11}(b) results in a trivial correction, 
$ih_\alpha e_{\alpha\beta\gamma}$, the second and the third diagrams 
[the term linear in $\Gamma(K)$ in eq.(\ref{enh})] give
$(i/4)h_\alpha e_{\alpha\beta\gamma} \partial \Sigma
/\partial\omega$, and the diagrams in Fig.~\ref{fig11}(c) 
[the term quadratic in $\Gamma(K)$ in the eq.(\ref{enh})] yield  
$-(5/4)ih_\alpha e_{\alpha\beta\gamma} \partial \Sigma/\partial\omega$.
Further separation of a quasiparticle contribution in a Green's 
function for physical operators $a_{{\bf k}\alpha}$ with polarization $\alpha$ taken in spiral coordinates,
$$
\frac{1}{\omega-\tilde \omega_{\bf k}-\Sigma({\bf k},\omega)+
\varepsilon \cdot h\cdot Z_{\bf k}}\approx \frac{Z_{\bf k}}
{\omega-\omega_{\bf k}+\varepsilon\cdot h},
$$
with $\varepsilon=-1,0,1$, results in a simple Zeeman splitting of 
the elementary triplet excitation.  Thus, instead of three degenerate 
triplets at $\omega_{\bf k}$ there are three excitations at 
$\omega_{\bf k}$ and $\omega_{\bf k}\pm h$, i.e., for magnetic moment 
$\mu=1$. We can now calculate the magnetic susceptibility at finite 
temperature. Since the temperature correction to the renormalization
constant is a very small, the main temperature correction to the 
self-energy is due to the second and third diagrams in Fig.~\ref{fig11}(b). 
We use the conventional Matsubara technique ($\omega\rightarrow i\omega_n$) 
to calculate this correction and obtain
\begin{equation}
\label{cor}
\delta\Sigma^{T,h}=-ih_{\alpha}e_{\alpha\beta
\gamma}\frac{\Gamma(p,0)}{4TL^2}\sum_{\bf q}
\frac{1}{\mbox{sinh}^2(\omega_{\bf q}/2T)}.
\end{equation}
Using (\ref{enh}) and (\ref{cor}) one can find the quasiparticle magnetic moment  
renormalized due to temperature:
\begin{equation}
\label{hef}
\mu({\bf p})=1-\frac{Z_{\bf p}\Gamma({\bf p},0)}{4TL^2}\sum_{\bf q}
\frac{1}{\mbox{sinh}^2(\omega_{\bf q}/2T)}.
\end{equation}
Then the uniform magnetic susceptibility per spin can be easily obtained 
from 
\begin{equation}
\label{chi}
\chi=\left.\frac{1}{2}\frac{\partial M}{\partial h}\right|_{h=0},\hspace{0.3cm}
M=-\frac{2}{N}\sum_p\frac{\partial n_{\bf p}}
{\partial \omega_{\bf p}}\mu({\bf p})\cdot h.
\end{equation}
Figure \ref{fig12} shows the  magnetic susceptibility as a function of 
temperature at $g=g_c$ and at $g=3.0$, as obtained from Eq.~(\ref{chi}) using 
the self-consistent solution of the Brueckner equations. In the same 
figure we also show our quantum Monte Carlo results. The agreement 
between the both methods is excellent at low temperature. The agreement at critical point is better than at $g=3.0$ because the value of triplet gap, $\Delta_0\approx 0.78J$, at $T=0,g=3.0$ in our analytic method is slightly different from numerical result $\Delta_0\approx 0.75$. This difference leads to a slightly different exponent for the magnetic susceptibility.

In the critical region, where $T\ll J,\delta g\ll g_c$, 
the main contribution to the integration in (\ref{hef}) is due to 
small momenta, $q\sim \Delta/\gamma$, thus the  susceptibility per 
spin can be written as
\begin{equation}
\chi=\frac{T}{2\pi \gamma^2}\Omega(x),
\end{equation}
with the universal function
\begin{equation}
\Omega(x)=\int_{y(x)}^{\infty}\frac{dx_1x_1 e^{x_1}}
{(e^{x_1}-1)^2}\left[1-\delta_0 T\int_{y(x)}^{\infty}\frac{x_1dx_1}{\mbox{sinh}
^2\frac{x_1}{2}}\right]=
\end{equation}
$$=\frac{1}{2}\left[y(x)\mbox{ctanh}\frac{y(x)}{2}-x\right]
\left[1-2\delta_0 T
\left(y(x)\mbox{ctanh}\frac{y(x)}{2}-x\right)
\right].
$$
The constant $\delta_0=Z_c\Gamma_c/(8\pi\gamma^2)\approx 0.06$ has been 
evaluated using the values of $\Gamma_c\approx 6.3J, 
Z_c\approx 0.8, \gamma\approx 1.9J$ taken from the
self-consistent solution of the Brueckner equations at zero 
temperature, see Refs.~\onlinecite{Sushkov,Shevchenko}. The 
universal function $\Omega(x)$ has the limiting behavior 
\begin{eqnarray}
&&\Omega(x)\approx\frac{\sqrt{5}}{2}y_0(1-\frac{2}{5}x)(1-2\delta_0 T
\sqrt{5}y_0),\hspace{0.3cm}
x\rightarrow 0\\
&&\Omega(x)\approx -\frac{x}{2}(1-\frac{2}{x})(1+2\delta_0\Delta_0),
\hspace{0.3cm}x\rightarrow -\infty\\
&&\Omega(x)\approx xe^{-x}(1+\frac{1}{x}), \hspace{0.3cm}x\rightarrow
\infty,
\end{eqnarray}
which is exactly the same as predicted using the nonlinear $\sigma$ 
model ($N\rightarrow \infty$) \cite{Sachdev0,Chubukov0} if corrections 
proportional to $\delta_0$ are neglected. These corrections are subleading
terms, but in the quantum critical and renormalized classical regimes 
they are parametrically larger then the subleading terms obtained for 
the $\sigma$ model ($N\rightarrow\infty$).

\section{Conclusions and Discussion}

In conclusion, we have presented an effective, self-consistent 
diagrammatic approach to describe the properties of the quantum 
double-layer Heisenberg antifferrromagnet at low temperature and 
$g\ge g_c$. To account for strong correlations between the elementary 
excitations (triplets), we applied the Brueckner approximation, treating 
the triplets as a dilute Bose gas with infinite on-site repulsion. 
We have also used a numerically exact quantum Monte Carlo simulation
method to obtain non-perturbative results for comparison. We have 
calculated temperature dependent spectra (including the triplet gap), 
dynamic and static structure factors, the specific heat, and the 
uniform magnetic susceptibility. The agreement between the analytical 
and numerical results is excellent.

Our analytical method involves approximations and an error of a few percent 
is always expected. The diagrammatic approach is valid for small on-site 
density of excitations because we neglected anomalous contributions in 
the vertex $\Gamma({\bf K})$. We emphasize, however, that the region of 
applicability of the present technique is surprisingly large. The reason 
for this can be traced to the fact that the density of triplets increases 
rather slowly as a function of temperature. We found that near the 
critical point: $\rho^T=0.15$ for $T=0.5J$ while $\rho=0.12$ 
at $T=0$. It justifies the approach even for a relatively high 
temperatures, $T\sim 0.5J$, when $g\ge g_c$. 

The Brueckner approach allows to calculate the elementary 
triplet spectrum $\omega_{\bf k}$ at finite temperature, not only for 
small momentum but in the whole Brillouine zone. Comparing the
analytical results for the dimensionless ratio 
$R({\bf q})=S({\bf q})/T\chi({\bf q})$ with quantum Monte Carlo 
data, we demonstrated that the dynamic structure factor is peaked at 
$\omega=\omega_{\bf k}$, the broadening of the triplet excitation 
is negligible and the elementary triplet is well defined for 
$g\ge g_c, T<0.5 J $.

In the $T\ll J$ and $g\rightarrow g_c$ limit, our analytical 
results show that correlation length, the specific heat, and the 
magnetic susceptibility are universal functions of the zero
temperature gap, $\Delta_0$, the spin-wave velocity, $\gamma$, and 
the temperature, $T$. The universal functions for the specific heat 
and the correlation length have the same limiting behavior as 
those predicted using the  nonlinear $\sigma$-model ($N\rightarrow\infty$),
while our universal function of the magnetic susceptibility contains
additional subleading terms. 

The agreement with the $\sigma$ model approach is due to the relative 
smallness of the quartic interaction, Eq.~(\ref{h4}), for the two-layer 
Heisenberg model considered in this paper. In this situation, the 
hard-core constraint is the most important and it is very natural 
that the results correspond to that of the $\sigma$ model. However, 
this situation is not general. There are many models where the 
quartic interaction is very important. It can even produce bound 
states of triplet spin-waves \cite{Sushkov1} which effectively 
change the number of relevant degree of freedom. In this situation, 
one can expect a very substantial deviation from the simple nonlinear
$\sigma$-model behavior, while the above Brueckner approach can still
be applied. An important example of such a system is the 2D $J_1-J_2$ model 
where the singlet bound state has an extremely low energy.\cite{Kotov}

The Brueckner theory describes quite well the quantum critical 
and quantum disordered phases near the quantum critical
point, as evidenced by the very good agreement with the quantum Monte
Carlo results. However, the description of the renormalized classical 
regime is not good in this theory because the ``Goldstone'' regime
with a Josephson length scale  $\xi_J$, where the spin dynamics is well 
described by rotationally averaged spin-wave fluctuations about a 
Neel-ordered ground state, is not present in our approach.
In the Brueckner approach the critical and Goldstone fluctuations are 
indistinguishable. It is well known, however, that there are only two 
gapless modes in the renormalized classical regime, while in our approach
there are three. The application of the Brueckner 
theory in this phase requires modifications; a nonzero sublattice 
magnetization $N_z$, i.e., a single particle condensate at zero 
momentum, has to be introduced. We plan to consider this in 
a future study.

The advantages of our analytic approach are that it is simple and 
captures the essential physics, as demonstrated here by comparing
results with quantum Monte Carlo simulations as well as with
predictions from the nonlinear $\sigma$ model. Obvious other applications 
of the Brueckner method include 2D Heisenberg models with frustration, 
Heisenberg ladders, as well one-dimensional spin chains at finite temperature. 

In addition to confirming the high accuracy of the Brueckner theory, the 
quantum Monte Carlo simulations performed here also show unambiguosuly that 
quantum critical behavior can be observed in the bilayer model at temperatures
as high as $T/J \alt 0.5$. We also obtained an improved
estimate of the critical coupling ratio; $g_c = 2.525 \pm 0.002$.

\section{Acknowledgments}

We are very grateful to V. Kotov, R. Singh, G. Gribakin and M. Kuchiev for stimulating 
discussions. This work was supported by a grant from the Australian Research 
Council. A. W. S. would like to thank the School of Physics at the 
University of New South Wales for hospitality and financial support 
during a visit. Support from the NSF under Grant No.~DMR-9712765 
is also acknowledged.

\vspace{0.5cm}

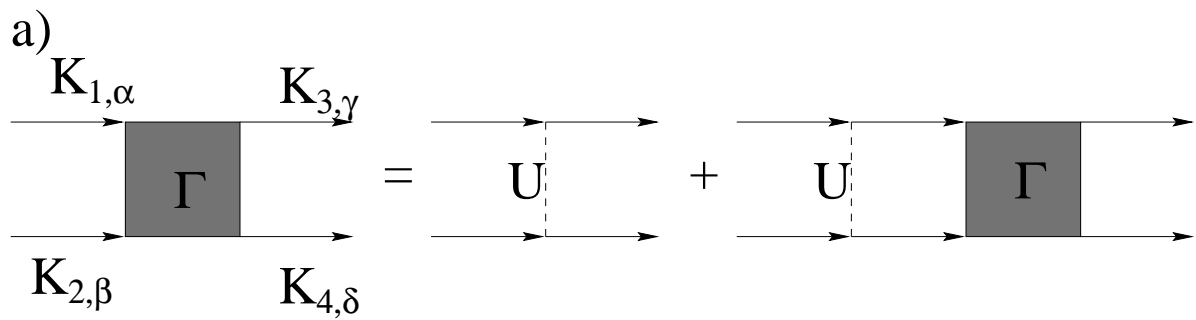
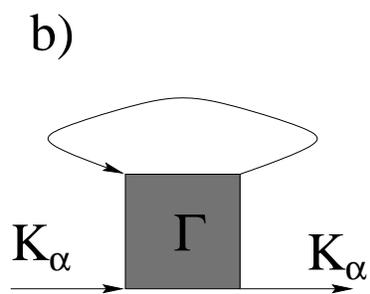
\begin{figure}
\caption{a) Equation for the scattering vertex $\Gamma({\bf K})$.
b) Diagram for the self-energy $\Sigma({\bf K})$ corresponding to $\Gamma({\bf K})$.} 
\label{fig1}
\end{figure}

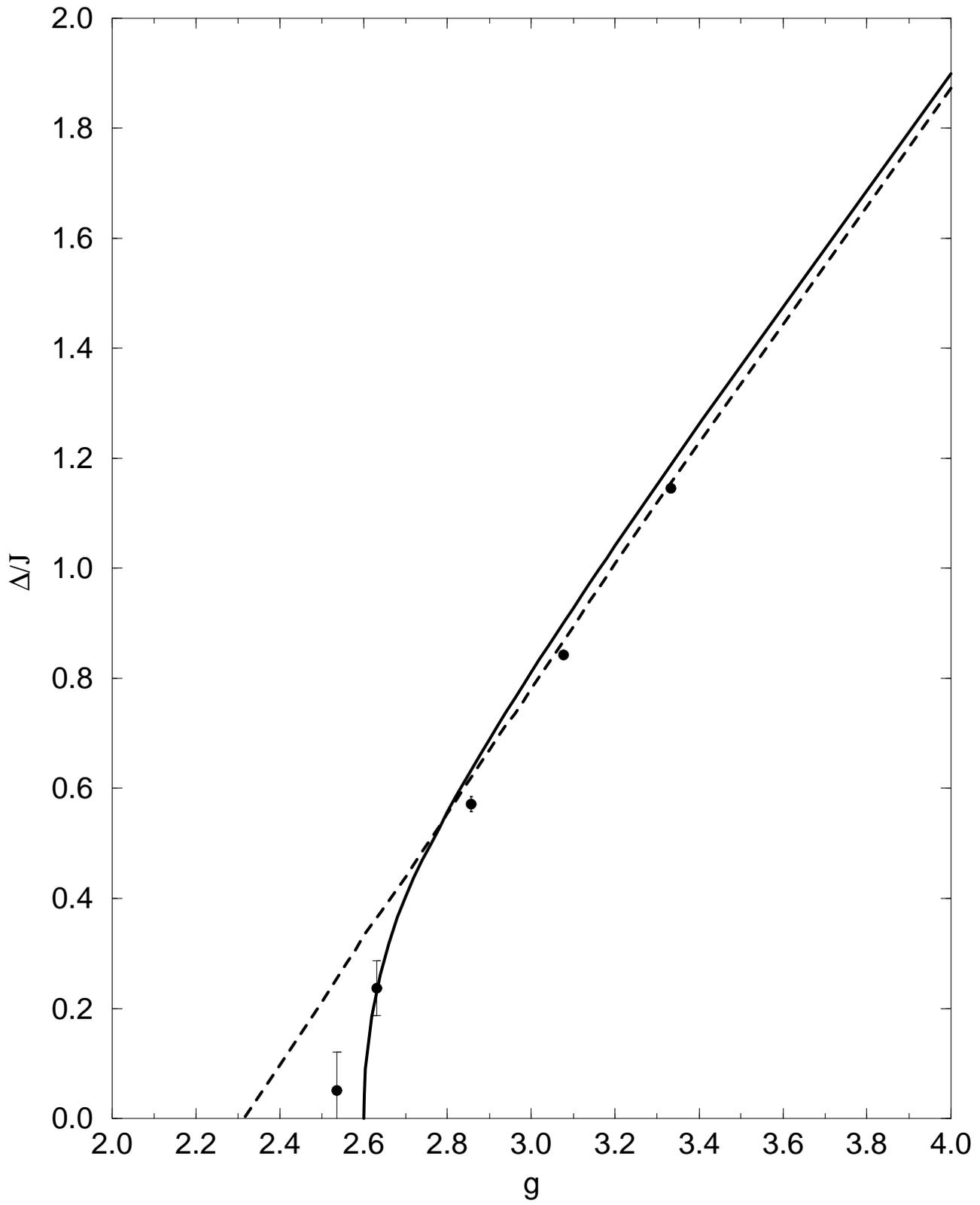
\begin{figure}
\caption{Triplet gap, $\Delta/J$,  as a function of coupling $g=J_\perp/J$. 
Solid and dashed lines are the results of the self-consistent solution 
using the Brueckner approach and the Bruekner equations linearized in 
density, respectively. The dots with error bars are estimates obtained 
by dimer series expansions.\protect\cite{Sushkov}}
\label{fig2}
\end{figure}

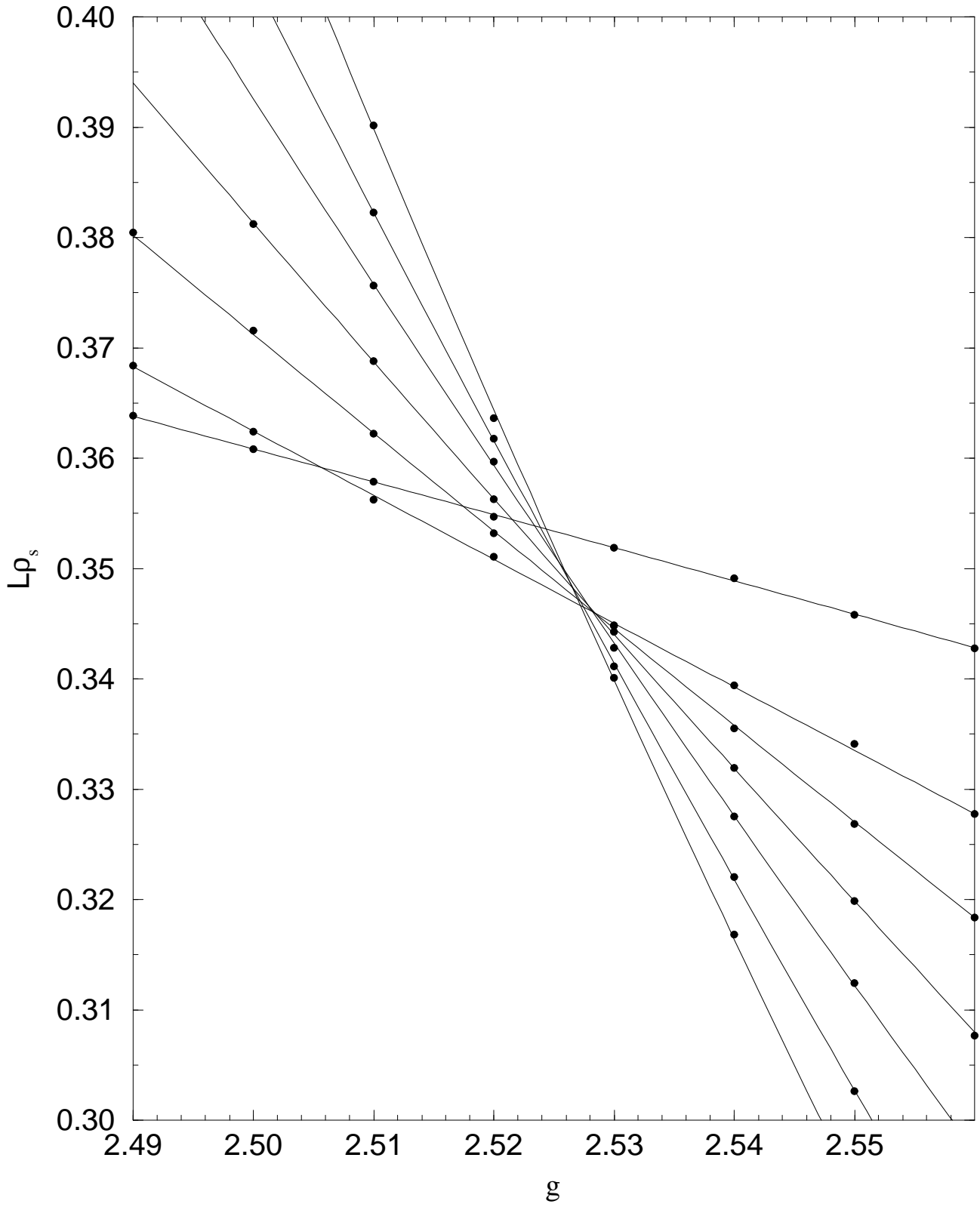
\begin{figure}
\caption{QMC results for the $T=0$ spin stiffness $\rho_s$ times the linear
system size $L$, for $L=4,6,\ldots,16,20$. The curves are quadratic 
fits to the QMC data (solid circles). The (negative) slope of
$L\rho_s$ vs $g$ increases with $L$. Statistical errors are typically
of the order of the radius of the circles.}
\label{fig3}
\end{figure}

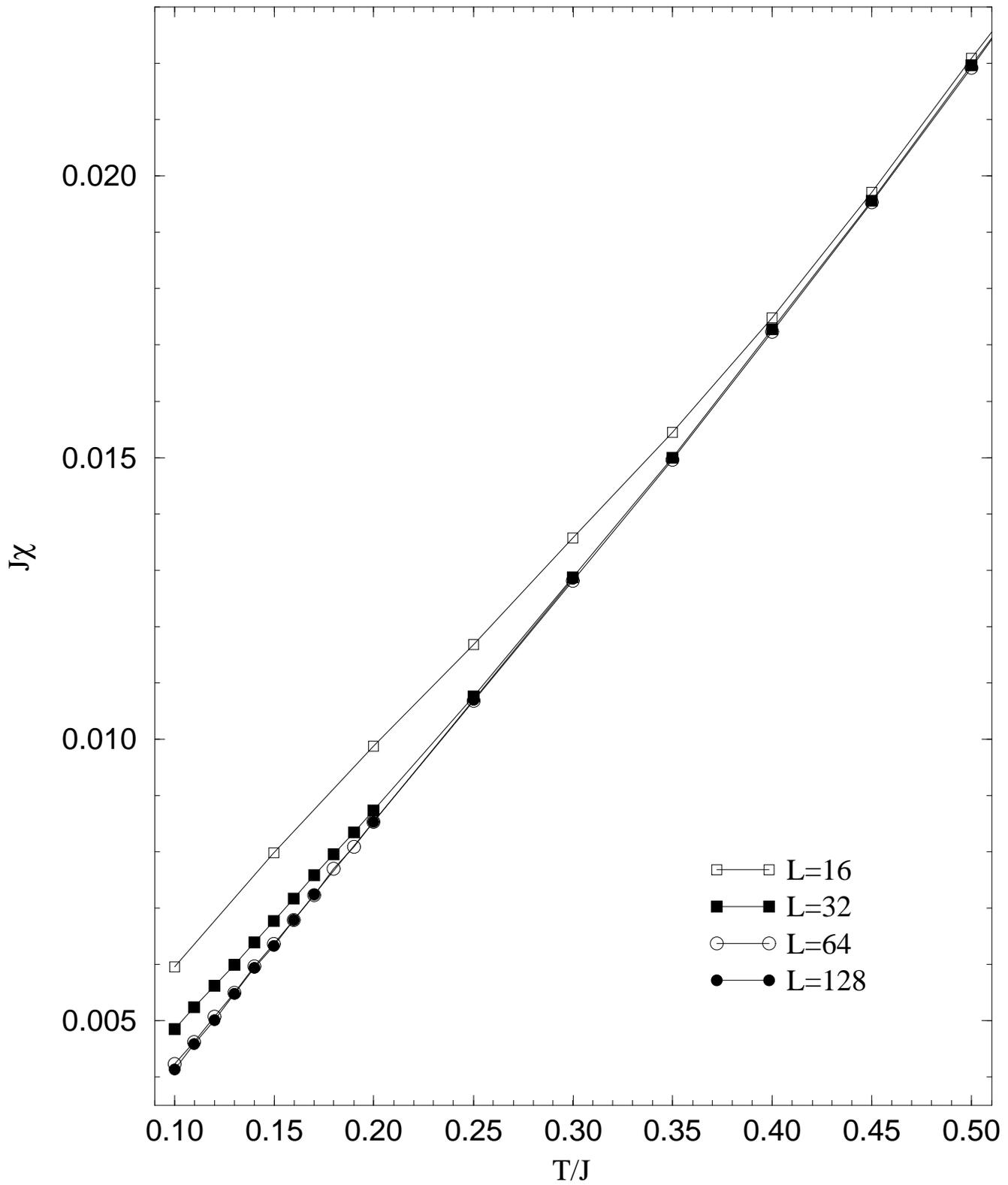
\begin{figure}
\caption{
The uniform magnetic susceptibility, $\chi J$, per spin vs 
temperature, $T/J$,  for different system sizes at a 
coupling $g=2.53$. Statistical errors are much smaller than the sympols.}
\label{fig4}
\end{figure}

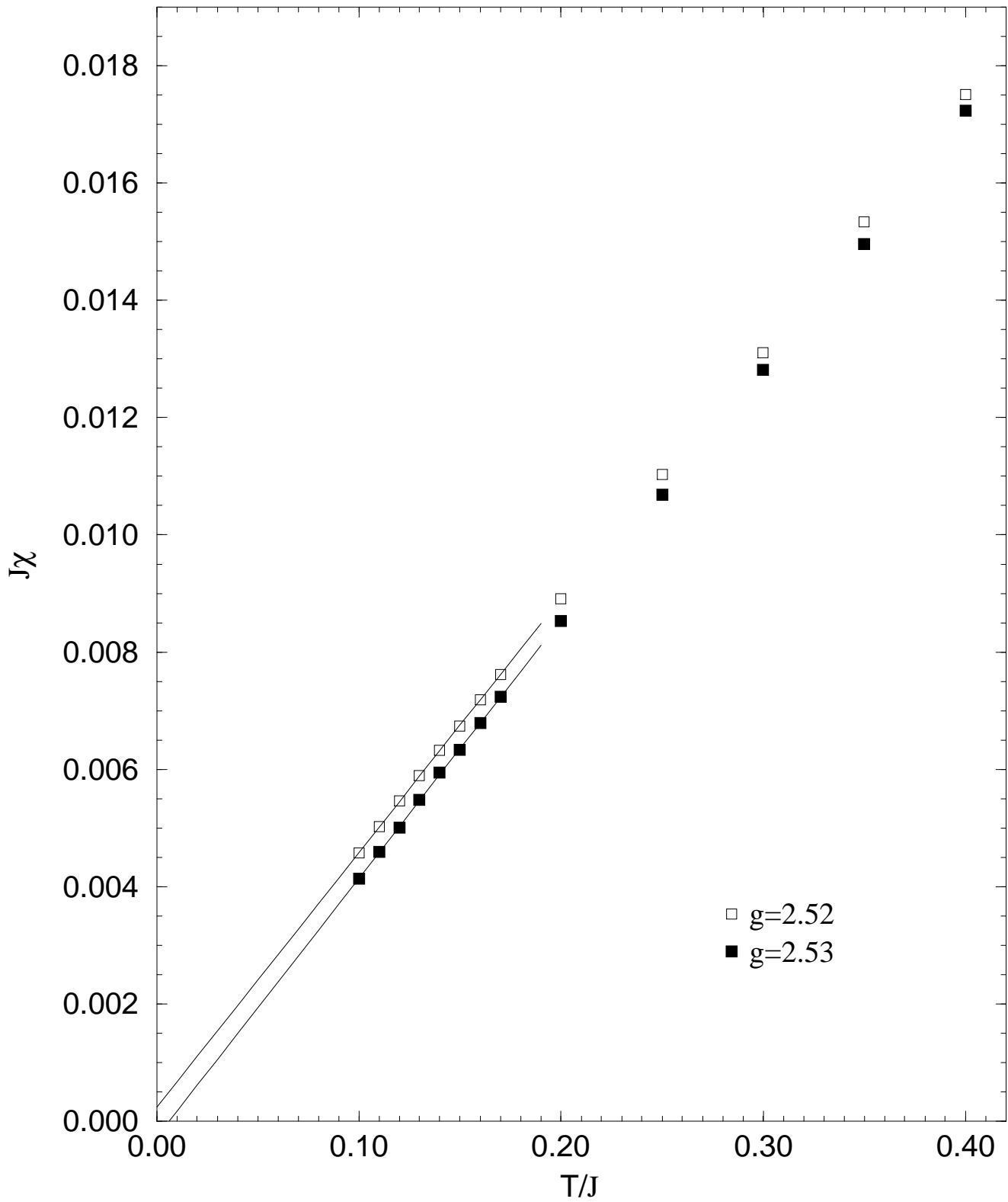
\begin{figure}
\caption{
The uniform magnetic susceptibility, $\chi J$, per spin vs 
temperature, $T/J$, for $g=2.52$ and $2.53$. The lines are
fits to the $T/J \le 0.17$ data.} 
\label{fig5}
\end{figure}

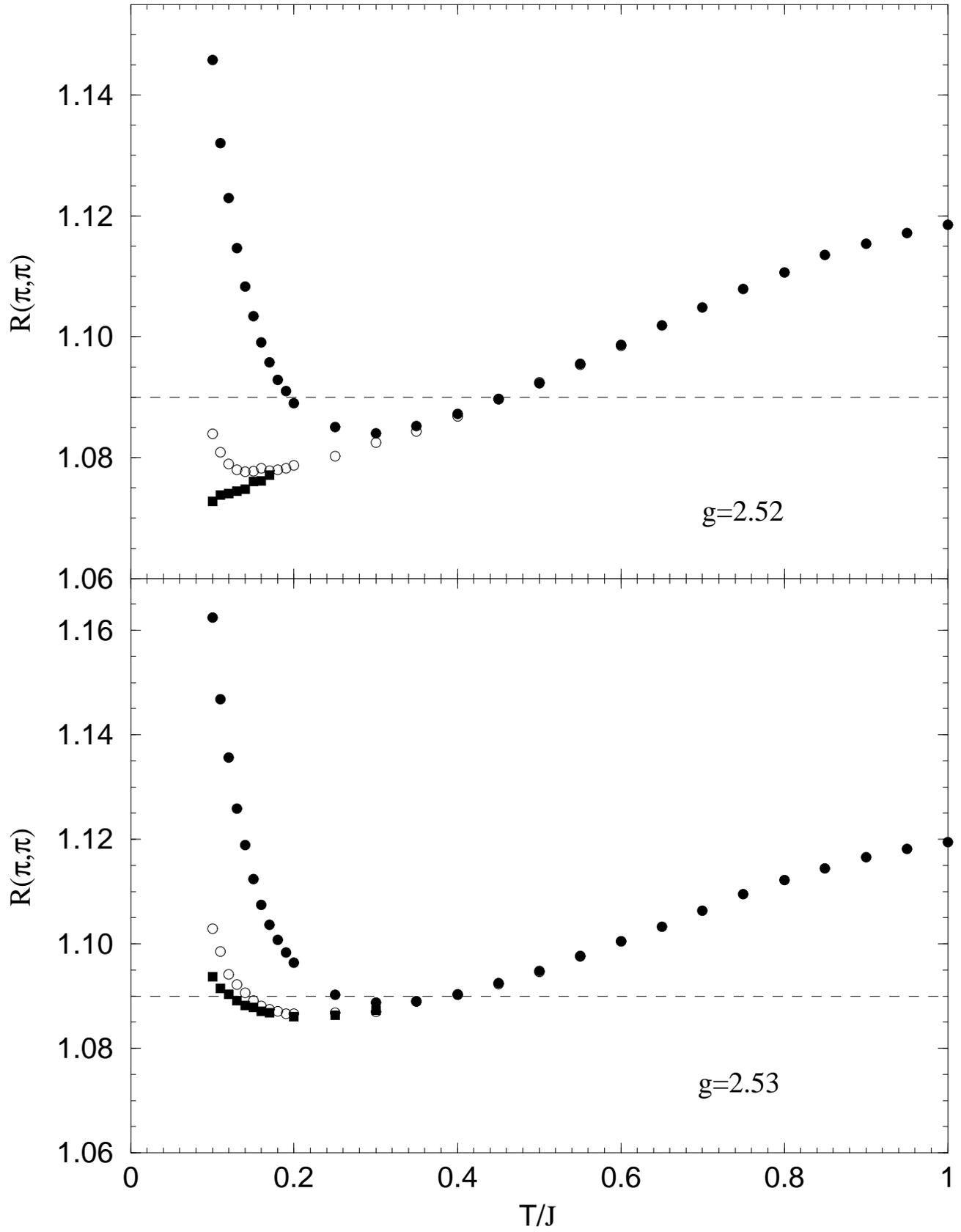
\begin{figure}
\caption{
The ratio of the staggered structure factor 
and $T$ times the staggered susceptibility for $L=32$ (solid circles), 
$L=64$ (open circles), and $L=128$ (solid squares), for a coupling 
slightly below ($g=2.52$) and above ($g=2.53$) the critical coupling 
as a function of temperature, $T/J$. Statistical errors are 
at most of the order of the size of the symbols. The dashed lines
indicate the constant value predicted within the nonlinear $\sigma$-model
approach.}
\label{fig6}
\end{figure}

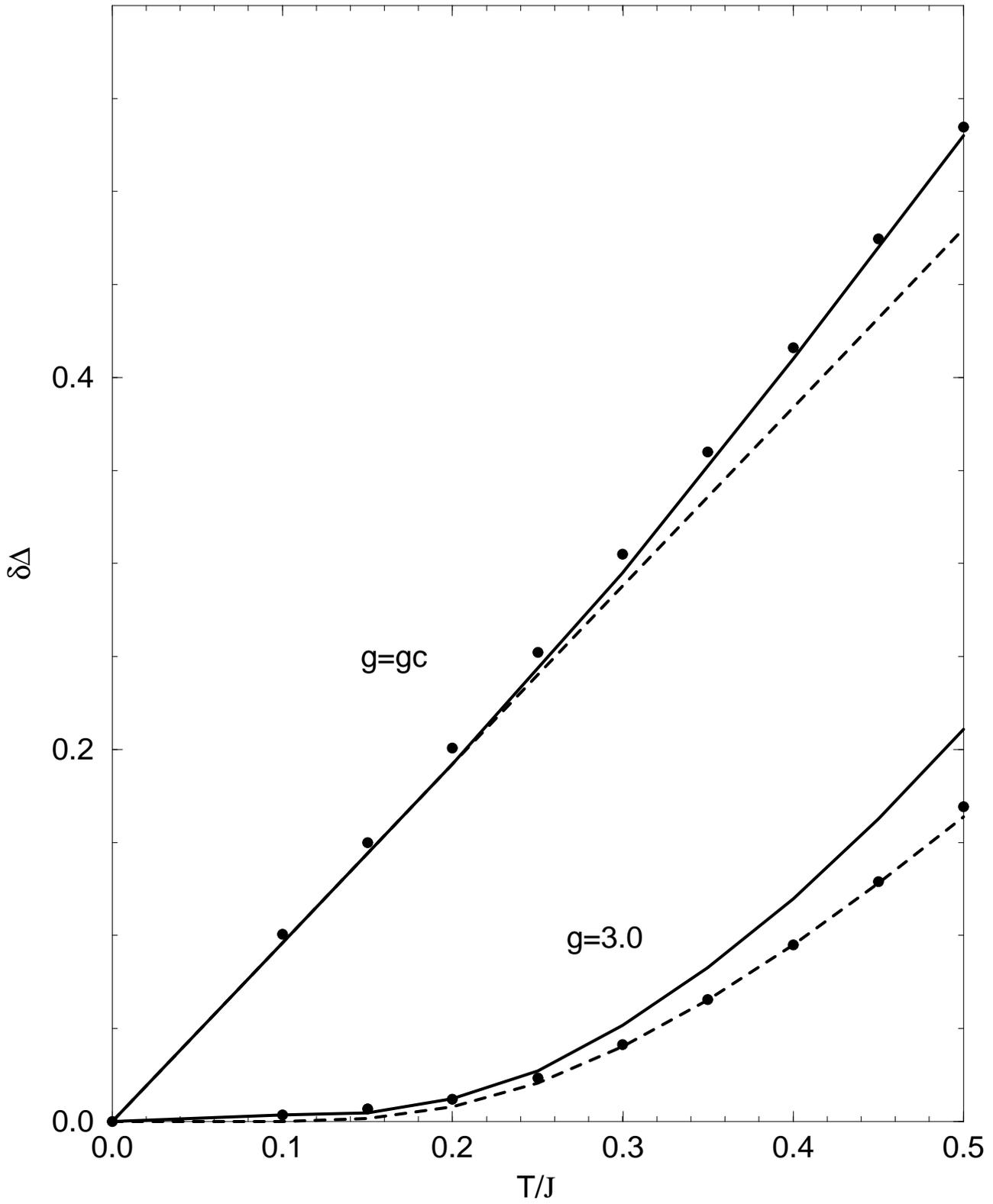
\begin{figure}
\caption{Deviation $\delta\Delta=(\Delta_T-\Delta_0)/J$ as a 
function of temperature, $T/J$, at the crititcal point, 
$g=g_c$, and in the quantum disordered regime at $g=3.0$. Solid lines 
are results of self-consistent solutions of the Brueckner equations 
linearized in density, dots are estimates extracted from Monte Carlo 
simulations for $R(0)$ using Eq.(\ref{wq0}), and dashed lines are 
predictions of the analytical Eq.~(\ref{a}).}
\label{fig7}
\end{figure}

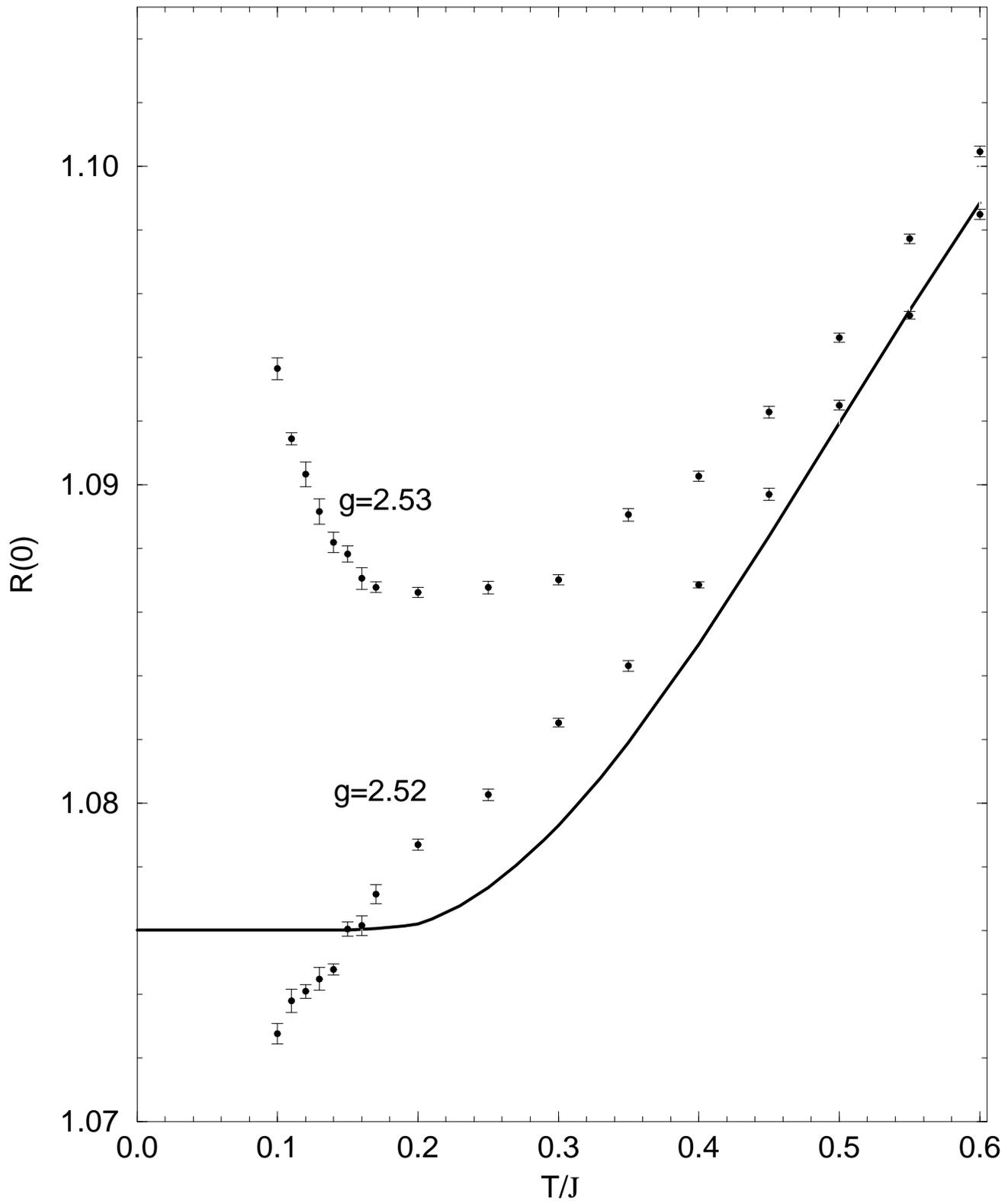
\begin{figure}
\caption{The dimensionless ratio $R(0)=S(0)/[T\chi(0)]$ as a function of 
temperature, $T/J$, for $g=2.52$ and $g=2.53$ obtained using 
quantum Monte Carlo simulation (dots). The solid line is the analytical 
result for $g=g_c$ obtained from the self-consistent solution of the
Brueckner equations.}
\label{fig8}
\end{figure}

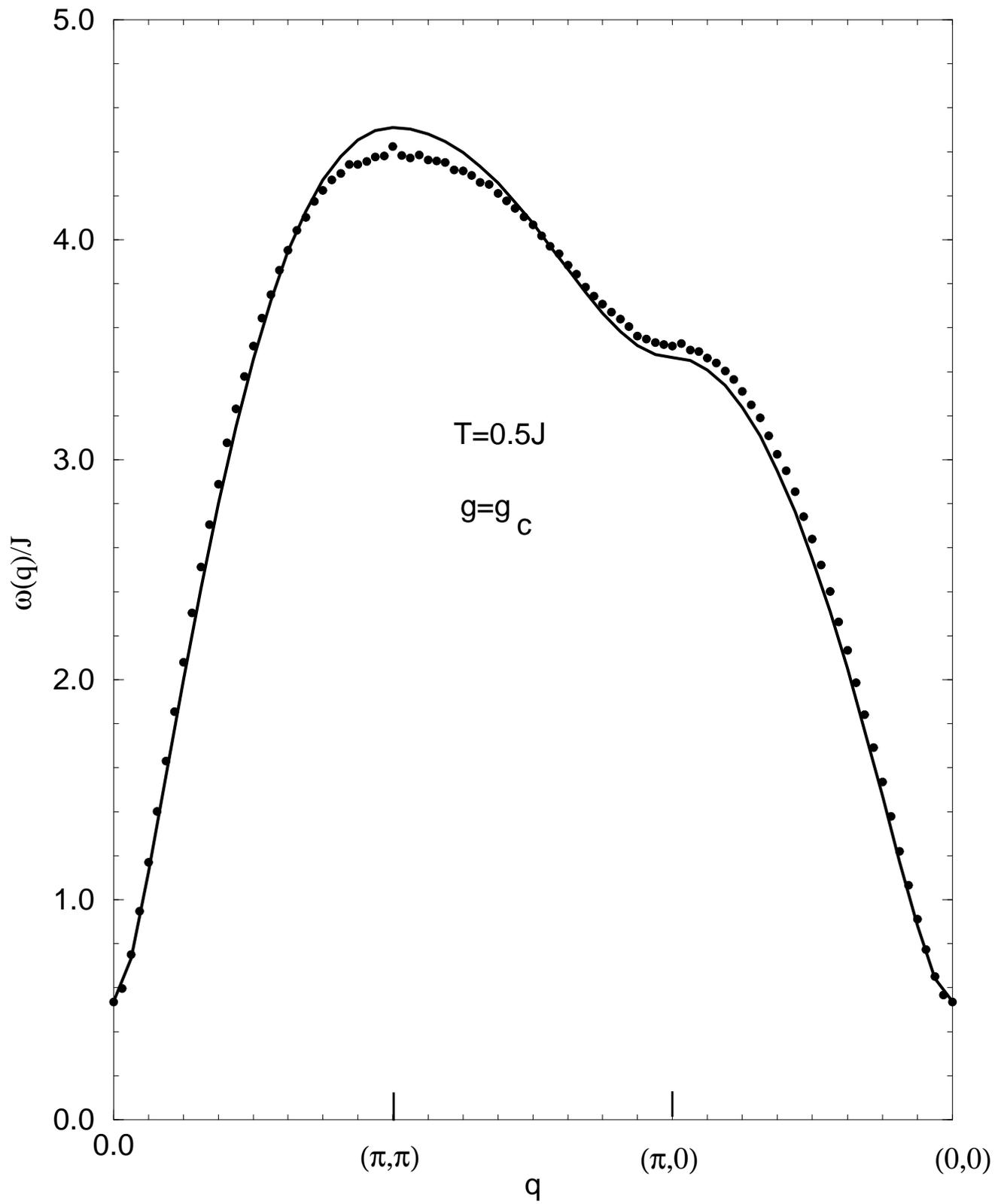
\begin{figure}
\caption{Elementary triplet spectrum, $\omega(q)/J$, along the
triangle in Brillouine zone, $q=(0,0)$$-(\pi,\pi)$$-(\pi,0)-(0,0)$, 
at $g=g_c, T=0.5J$. The solid line is the result of a 
self-consistent solution of the Brueckner equations, the dots are
estimates extracted from Eq.~(\ref{wq}) and Monte Carlo results 
for $R(q)$.} 
\label{fig9}
\end{figure}

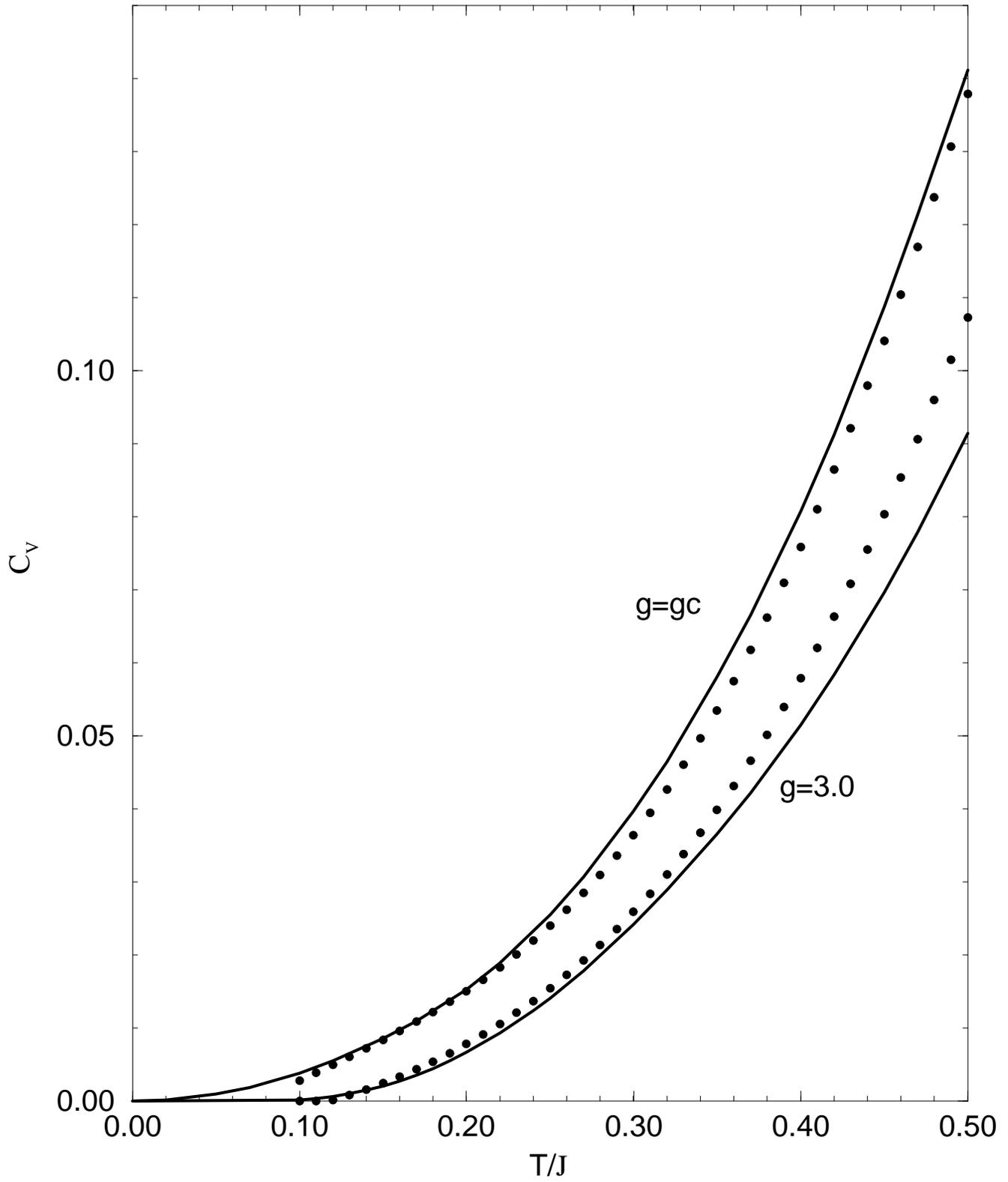
\begin{figure}
\caption{Specific heat per spin, $C_V$, as a function of temperature 
$T/J$ at critical point $g=g_c$ and $g=3.0$. Solid lines are  
the results of self-consistent solution of Brueckner equations, 
dots are estimates obtained by Monte Carlo simulations.}
\label{fig10}
\end{figure}

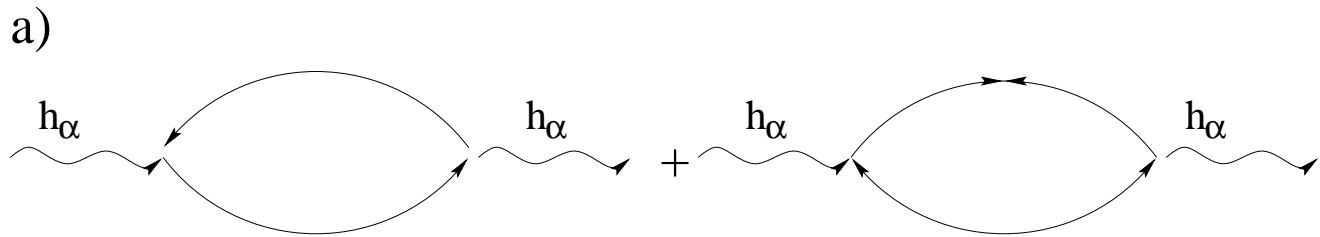
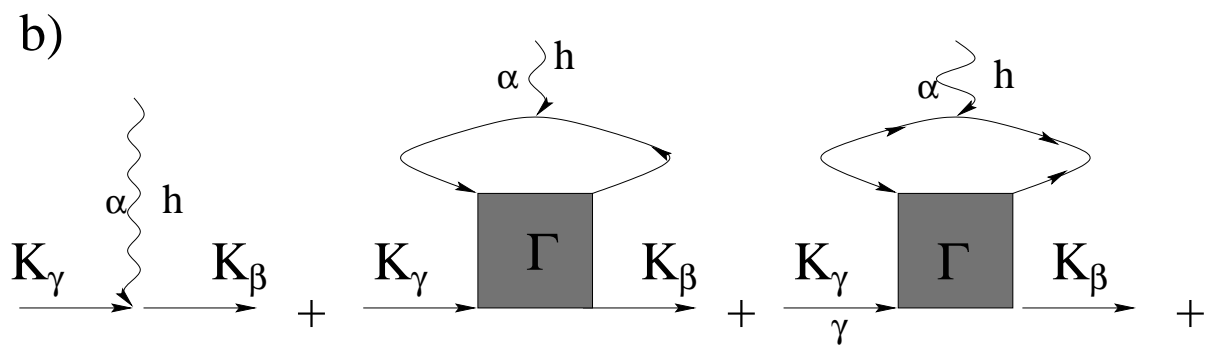
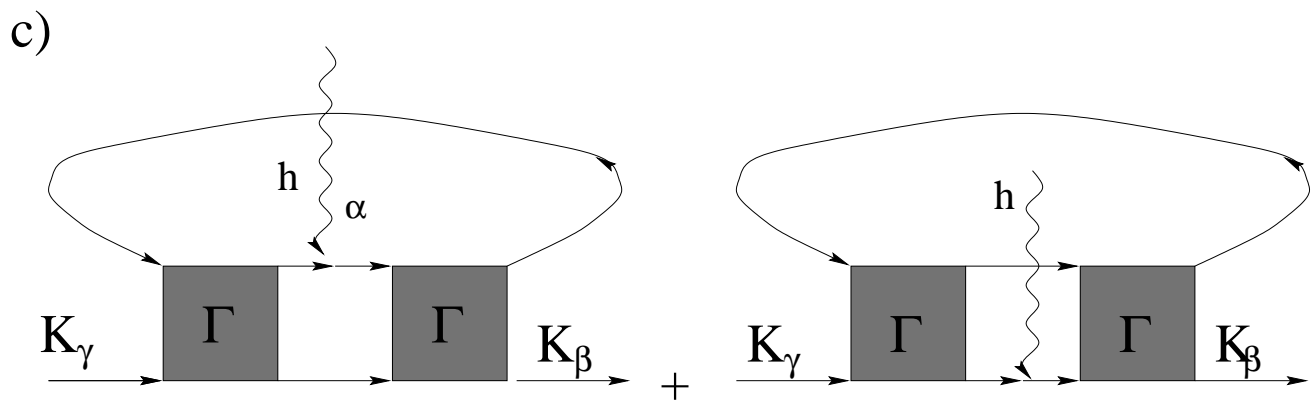
\begin{figure}
\caption{a) Diagrams for the second order correction to the total 
energy due to a magnetic field perturbation. b),c) the diagrams for the 
correction to the normal self energy due magnetic field.}
\label{fig11}
\end{figure}

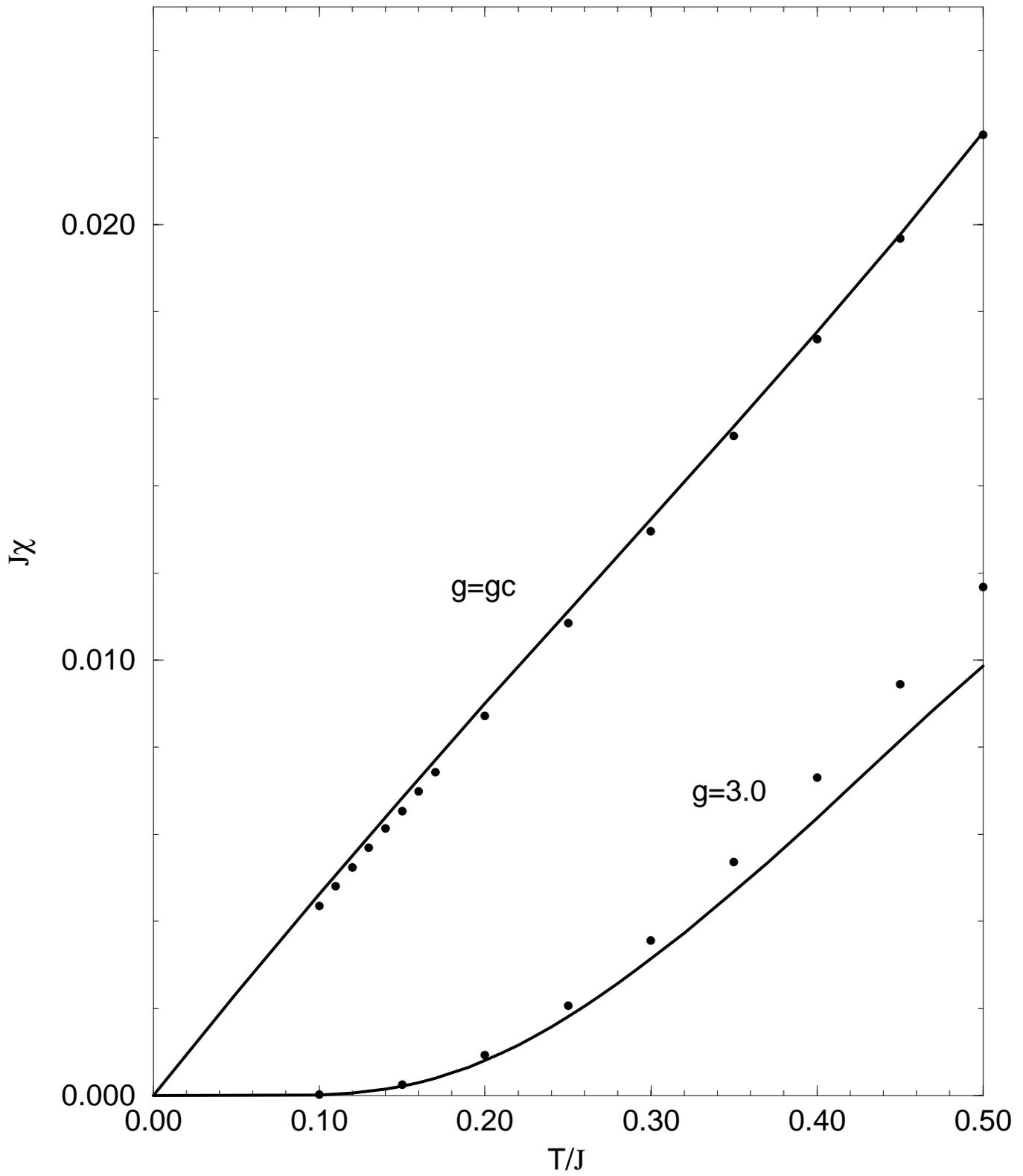
\begin{figure}
\caption{
Magnetic susceptibility per spin, $\chi J$, as a function of temperature, $T/J$, at 
$g=g_c$ and $g=3.0$. The solid lines are results of self-consistent 
solutions of the Brueckner equations linearized in density. The dots 
are Monte Carlo results.}
\label{fig12}
\end{figure}

\end{document}